\documentclass[a4paper,10pt]{article}

\usepackage{amsfonts}
\usepackage{amsmath}
\usepackage[all]{xy}
\usepackage{verbatim}
\usepackage{rotating}

\usepackage[ps2pdf]{hyperref}
\hypersetup{pdftitle = {Formula for Fixed Point Resolution Matrix of Permutation Orbifolds}}
\hypersetup{pdfauthor = {Michele Maio, Bert Schellekens}}
\hypersetup{pdfsubject = {CFT-String Theory}}
\hypersetup{pdfkeywords = {Orbifolds, Extensions, Fixed Point Resolution}}

\numberwithin{equation}{section}

\begin{document}

\title{Formula for Fixed Point Resolution Matrix of Permutation Orbifolds}
\author{M. Maio$^{1}$ and A.N. Schellekens$^{1,2,3}$\\~\\~\\
\\
$^1$Nikhef Theory Group, Amsterdam, The Netherlands
\\
~\\
$^2$IMAPP, Radboud Universiteit Nijmegen, The Netherlands
\\
~\\
$^3$Instituto de F\'\i sica Fundamental, CSIC, Madrid, Spain}

\maketitle

\begin{abstract}
We find a formula for the resolution of fixed points in extensions of permutation orbifold conformal field theories by its (half-)integer spin simple currents. We show that the formula gives a unitary and modular invariant $S$ matrix.
\end{abstract}

Preprint: NIKHEF 2009-024

\clearpage
\tableofcontents

\section{Introduction}
In a series of recent papers \cite{Maio:2009kb,Maio:2009cy} we have started to study the problem of resolving the fixed points 
\cite{Fuchs:1996dd,Schellekens:1999yg,Schellekens:1989uf} in simple current \cite{Schellekens:1989am,Intriligator:1989zw,Schellekens:1990xy,Schellekens:1989dq} extensions of permutation orbifold \cite{Klemm:1990df,Borisov:1997nc} conformal field theories \cite{Belavin:1984vu}. The aim of this paper is to give a general solution to this problem, going much beyond the specific examples discussed previously.

Simple currents $J$ are special fields: those with simple fusion rules with any other field in the theory. They are very important ingredients of a CFT, since they allow to modify the theory in a well-controlled way by projecting out some fields and re-organizing the remaining into new ones. In practice, what one does is: compute the monodromy charge of any field $i$ with respect to (w.r.t.) $J$, $Q_J(i)$; project out those fields which have non-integer monodromy charge; organize the surviving fields into orbits under $J$. 

In this paper we will mostly look at order-two simple currents, i.e. those with $J^2=1$, for which the $J-$orbits can have length equal to one or two at most. The normal (and easy to handle) fields are those with length two: $(i,J\cdot i)$. More special are those orbits with length one: $J\cdot f=f$. A field $f$ satisfying this property is called a \textit{fixed point} of the current $J$. Fixed points can arise only when the current $J$ has integer or half-integer spin, given by its weight $h_J$. In the extension, they give rise to more than one field, whose number is equal to the order of the current. In this paper then every fixed points will split into two fields in the extended theory.

In any CFT, two of the most important objects are the modular $S$ and $T$ matrices. $T$ is a diagonal matrix of phases and contains information about the weights of the fields in the theory; $S$ is symmetric and unitary and allows to compute the fusion coefficients (conceptually analogous to the Clebsch-Gordan series) between two representations via the Verlinde formula \cite{Verlinde:1988sn}. Without fixed points, there is no difficulty in deriving the $S$ matrix of the extended theory, sometimes denoted by $\tilde{S}$, from the $S$ matrix of the unextended CFT. On the contrary, when fixed points are present, not only the extended $S$ matrix is problematic to derive but also it is affected by an intrinsic ambiguity related to the freedom that we have in choosing the order of the splitted fields coming from the same fixed point. This issue has already been addressed in the past \cite{Fuchs:1996dd} and the outcome was that we can write the matrix $\tilde{S}$ in terms of a set of matrices $S^J$, one for each simple current $J$, and hence the problem of resolving the fixed points is re-formulated as the problem of determining those $S^J$ matrices.

Using the formalism developed in \cite{Fuchs:1996dd}, we can trade our ignorance about $\tilde{S}$ with a set of matrices $S^J$, one for every simple current $J$, according to the formula
\begin{equation}
\label{main formula for f.p. resolution}
\tilde{S}_{(a,i)(b,j)}=\frac{|G|}{\sqrt{|U_a||S_a||U_b||S_b|}}\sum_{J\in G}\Psi_i(J) S^J_{ab} \Psi_j(J)^{\star}\,,
\end{equation}
These $S^J_{ab}$'s are non-zero only if both $a$ and $b$ are fixed points. This equation can be viewed as a Fourier transform and the $S^J$'s as Fourier coefficients of $\tilde{S}$. The prefactor is a group theoretical factor acting as a normalization and the $\Psi_i(J)$'s are the group characters acting as phases. Unitarity and modular invariance of $\tilde{S}$ implies unitarity and modular invariance of the $S^J$'s \cite{Fuchs:1996dd}:
\begin{equation}
\label{modular}
S^J\cdot (S^J)^\dagger=1 \qquad (S^J\cdot T^J)^3=(S^J)^2\,.
\end{equation}
In this way, the problem of finding $\tilde{S}$ is equivalent to the problem of finding the set of matrices $S^J$. 

The unextended theory that we can consider before the extension can be any CFT $\mathcal{A}$. It can be also a tensor product of different CFT's, $\mathcal{A}_1\otimes\dots\otimes\mathcal{A}_n$, or even a coset theory of the form $G/H$. All these cases have already been considered in the past and their $S^J$ matrices are known by now. In fact, these matrices have been found for all WZW models \cite{Fuchs:1996dd,Fuchs:1995zr}, their simple current extensions \cite{Schellekens:1999yg} and for coset conformal field theories \cite{Schellekens:1989uf}. In this paper we will consider the permutation orbifold as the unextended CFT, for which the $S^J$ matrices are in general not known yet. We restrict ourselves to the case of $\mathbb{Z}_2$ orbifolds \cite{Klemm:1990df,Borisov:1997nc}, where we mod out by the cyclic permutation that exchanges the two factors:
\begin{equation}
 \mathcal{A}_{\rm perm} \equiv \mathcal{A}\times \mathcal{A}/\mathbb{Z}_2\,.
\end{equation}
Larger orbifolds would be possible \cite{Bantay:1997ek,Bantay:1999us}, but they are much more involved and we will not treat them here.

The matrices $S^J$ are restricted not only by modular invariance and unitarity, but also by the condition that the full matrix 
$\tilde{S}_{(a,i)(b,j)}$ acts on a set of characters with positive integer coefficients, that the Verlinde formula yields non-negative
integer coefficients and that there is a corresponding set of fusing and braiding matrices that satisfy all hexagon and pentagon
identities. In other words, all the usual conditions of rational conformal
field theory should be  satisfied. However, all these additional constraints are very hard to check, and modular invariance and unitarity
are very restrictive already. Experience so far suggests that for generic formulas ({\it i.e.} formulas valid for an entire class, as opposed
to special solutions valid only for a single RCFT) this is sufficient. We do not know any general results concerning the uniqueness
of the solutions to (\ref{modular}), but there is at least one obvious, and irrelevant ambiguity. If $S^J$ satisfies (\ref{modular}), clearly
$U^{\dagger} S^J U$ satisfies it for any unitary matrix $U$ that commutes with $T$. Since we are aiming for a generic solution, we
may assume that $T$ is non-degenerate; accidental degeneracies in specific cases cannot affect a generic formula. This reduces $U$
to a diagonal matrix of phases. The matrix $\tilde{S}_{(a,i)(b,j)}$ must be symmetric, and this has implications for the symmetry of the
matrix $S^J$. In particular, if $J$ is of order 2 (the case considered here), the matrix $S^J$ must be 
symmetric itself \cite{Fuchs:1996dd}. This requirement reduces $U$ to a diagonal matrix of signs. These signs are irrelevant:
they simply correspond to a relabeling of the two components of each resolved fixed point field. Note that the matrix $S$ itself
also satisfies (\ref{modular}), but here there is no such ambiguity: $S$ acts on positive characters, and any non-trivial sign choice
would affect the positivity of $S_{0i}$. However, $S^J$ acts on {\it differences} of characters, and hence satisfies no such 
restrictions. 

For WZW models it was possible to obtain an explicit character representation of $S^J$ in terms of so-called
``twining characters" \cite{Fuchs:1995zr}. In the present case, however, we resort to the strategy of obtaining an {\it ansatz}
for $S^J$ based on its modular properties, along the lines of \cite{Schellekens:1989uf}. To arrive at this {\it ansatz} we make
use of the following pieces of information:
\begin{itemize}
\item{The matrix $S$ of the unextended $\mathbb{Z}_2$ orbifold, derived in \cite{Borisov:1997nc}. This is the matrix $S^J$ for the special case $J=0$, which fixes all fields in the CFT. We will denote it by $S^{BHS}$.}
\item{The matrix $S^J$ for the anti-symmetric component of the identity \cite{Maio:2009kb}. 
This matrix could be derived because this simple current
undoes the permutation orbifold and gives back the original tensor product.}
\item{The matrix $S^J$ for some cases where $J$ has spin 1 \cite{Maio:2009kb}. 
Here we used the fact that the simple current extension can be identified with a known WZW model. This allowed us
to determine $S^J$ for the vector current of $SO(N)$ level 1.}
\item{Using triality in $SO(8)$  this could be generalized to the spinor currents of $SO(8)$ level 1, and 
from there to all spinor currents of $SO(2n)$ level 1 \cite{Maio:2009cy}, which have very similar modular properties.}
\end{itemize}

Here we will use these previous works as ``stepping stones" towards a general ansatz, which includes all
the aforementioned results as special cases, and has a far larger range of validity. In particular, the results
of our foregoing work 
\cite{Maio:2009kb,Maio:2009cy}, were limited to low levels, such as in the permutation orbifold of $B(n)_1$, $D(n)_1$ and $A(1)_k$ (completely for $k=2$ and $k$ odd, partially for $k$ even). By an educated guess, one could very well suspect that this formula would depend on a few quantities of the original or mother CFT $\mathcal{A}$, such as its $S$ matrix, its $P$ matrix, the weight $h_J$ of the simple current $J$, etc. This is the problem that we address and solve in this paper. The formula which we obtain is valid
for any order two simple current $J$ of any order two permutation orbifold. In particular, this extends the foregoing results
for $B(n)$, $D(2n)$ and $A(1)$ to arbitrary level, but it also includes permutation orbifolds of many other WZW models such as $C(n)$, 
$E(7)$, as well as the permutation orbifolds of 
many coset CFT's, such as the $N=0$ and $N=1$ minimal superconformal models and some of the currents of the $N=2$ minimal superconformal
models.   
Not included are fixed points of simple currents of orders larger than two, which occur for example in 
the permutation orbifolds
of $A(2)$ level $3k$, or $D(2n+1)$ for even level.

The plan of this paper is as follows.\\
In order to make this paper as self-contained as possible and to fix our notation, we start by reviewing the construction of the permutation orbifold, its $S^{BHS}$ matrix \cite{Borisov:1997nc}, together with its simple current and fixed point structure \cite{Maio:2009kb}.\\
In section \ref{section_simple_ansatz}, we give an ansatz for the simplified case when the mother theory has no fixed points. This restriction is suggested by the fact that some orbifold fixed points are not present, hence it is much easier to guess the ansatz. We prove that the simplified ansatz is unitary and modular invariant.\\
In section \ref{section_ansatz}, we extend the ansatz to all cases and comment about its unitarity and modular invariance.\\
The complete proof that our ansatz is actually unitary and modular invariant is relegated to the appendix.

\section{The permutation orbifold}
In this section we review a few facts that will be relevant about permutation orbifolds. A $\mathbb{Z}_2-$permutation orbifold of a given CFT $\mathcal{A}$ is defined as:
\begin{equation}
 \mathcal{A}_{\rm perm} \equiv \mathcal{A}\times \mathcal{A}/\mathbb{Z}_2\,.
\end{equation}
Moding out by $\mathbb{Z}_2$ means that the spectrum must contain fields that are symmetric under the interchange of the two factors. This theory admits an untwisted and a twisted sector. The untwisted fields are those combinations of the original tensor product fields that are invariant under this flipping symmetry. Their weights are simply given by the sum of the two weights of each single factor. There are two kinds of untwisted fields:
\begin{itemize}
\item \textit{diagonal}, denoted by $(i,\chi)$, with $\chi=0,\,1$, corresponding to the combination $\phi_i\otimes\phi'_i + (-1)^\chi \phi'_i\otimes\phi_i$, where $\phi'_i$ denotes the first non-vanishing descendant of the $\mathcal{A}-$field $\phi_i$, (corresponding to symmetric and anti-symmetric representations), with weight
\begin{equation}
h_{(i,\psi)}=2\,h_i +\delta_{i,0}\cdot\delta_{\psi,1}\,;
\end{equation}
\item \textit{off-diagonal}, denoted by $(mn)$, with $m<n$, corresponding to the combination $\phi_m\otimes\phi_n + \phi_n\otimes\phi_m$, with weight
\begin{equation}
h_{(mn)}=h_m+h_n\,.
\end{equation}
\end{itemize}
Twisted fields are required by modular invariance \cite{Klemm:1990df}. In general, for any field $\phi_i$ in $\mathcal{A}$, there are two twisted fields in the orbifold theory, labelled by $\chi=0,1$. We denote twisted fields by $\widehat{(i,\chi)}$. Their weights were derived in \cite{Klemm:1990df} and are given by
\begin{equation}
\label{KS weight for twisted fields}
 h_{\widehat{(i,\chi)}}=\frac{h_i}{2}+\frac{c}{24}\frac{(\lambda^2-1)}{\lambda}+\frac{\chi}{2}\,,
\end{equation}
where $h_i\equiv h_{\phi_i}$ and $c$ is the central charge of $\mathcal{A}$. Here, $\lambda=2$.

The $S$ matrix of $\mathcal{A}_{\rm perm}$ is known from \cite{Borisov:1997nc}. We will call it $S^{BHS}$ and it is given by
\begin{subequations}
\label{BHS}
\begin{eqnarray}
S_{(mn)(pq)}&=&S_{mp}\,S_{nq}+S_{mq}\,S_{np} \\
S_{(mn)\widehat{(p,\chi)}}&=&0 \\
S_{\widehat{(p,\phi)}\widehat{(q,\chi)}}&=&\frac{1}{2}\,e^{2\pi i(\phi+\chi)/2} \,P_{ip} \\
S_{(i,\phi)(j,\chi)}&=&\frac{1}{2}\,S_{ij}\,S_{ij} \\
S_{(i,\phi)(mn)}&=&S_{im}\,S_{in} \\
S_{(i,\phi)\widehat{(p,\chi)}}&=&\frac{1}{2}\,e^{2\pi i\phi/2} \,S_{ip} \,,
\end{eqnarray}
\end{subequations}
where the $P$ matrix (introduced in \cite{Bianchi:1990yu}) is defined by $P=\sqrt{T}ST^2S\sqrt{T}$.

If there is any integer or half-integer spin simple current in $\mathcal{A}$, it gives rise to an integer spin simple current in $\mathcal{A}_{\rm perm}$, which can be used to extend the orbifold CFT itself. We can denote the extended permutation orbifold by $\tilde{\mathcal{A}}_{\rm perm}$. In the extension, some fields are projected out while the remaining organize themselves into orbits of the current. Typically untwisted and twisted fields do not mix among themselves. As far as the new spectrum is concerned, we do know that these orbits become the new fields of $\tilde{\mathcal{A}}_{\rm perm}$, but we do not normally know the new $S$ matrix, $\tilde{S}$.

As already mentioned, the problem of finding $\tilde{S}$ is equivalent to the problem of finding the set of matrices $S^J$, one for each simple current $J$. As a starting point, it will be useful to know what are the simple currents arising in the extended permutation orbifold. From the sufficient and necessary condition $S^{BHS}_{J0}=S^{BHS}_{00}$ \cite{Dijkgraaf:1988tf}, it is straightforward to discover that they correspond to the symmetric ($\psi=0$) and anti-symmetric ($\psi=1$) representations (hence diagonal fields) of the simple currents in the mother theory $\mathcal{A}$ \cite{Maio:2009kb}. It will then make sense to denote simple currents of the permutation orbifolds as $(J,\psi)$, being $J$ the corresponding simple current in the mother theory. There are no other possible combinations of $\mathcal{A}-$fields that become simple currents in the orbifold. Hence, one simple current in $\mathcal{A}$ generates two simple currents in $\mathcal{A}_{\rm perm}$.

Another useful piece of information is the fixed point structure arising in $\mathcal{A}_{\rm perm}$. By studying the fusion coefficients one can show that \cite{Maio:2009kb}:
\begin{itemize}
\item diagonal fields: $(i,\phi)$ is a fixed point of $(J,\psi)$ if  $\psi=0$ and if $i$ is a fixed  point of $J$, i.e. $Ji=i$;
\item off-diagonal fields: $(mn)$ is a fixed point of $(J,\psi)$ 
\begin{itemize}
\item either if $m$ and $n$ are both fixed points of $J$, i.e. $Jm=m$ and $Jn=n$,
\item or if $m$ and $n$ are in the same $J-$orbit, i.e. $Jm=n$;
\end{itemize}
\item twisted fields: $\widehat{(p,\phi)}$ is a fixed point of $(J,\psi)$ if $Q_J(p)=\frac{\psi}{2}+2\,h_J\,\,{\rm mod}\,\, \mathbb{Z}$, independently of $\phi$.
\end{itemize}
For the twisted fixed points, see the proof\footnote{In \cite{Maio:2009kb} we considered only (half-)integer $h_J$. In this paper we will have to look at currents with $h_J \in \frac{1}{4}\,\mathbb{Z}_{\rm odd}$ as well, since they give rise to half-integer spin currents in the orbifold and those can have fixed points.} in the appendix, formula (\ref{twisted fp's}). Observe that for (half-)integer spin simple currents we can drop the additional $2h_J$ from the monodromy charge.

Also note that there exist diagonal fixed points only for the symmetric representation of the simple current and that the twisted fixed points are determined by $Q_J(p)$, the monodromy charge of $p$ w.r.t. $J$. Moreover, we will often have to distinguish between the two types of fixed points coming from the off-diagonal sector: for obvious reasons, we will call them \textit{fixed-point-like off-diagonal fields} and \textit{orbit-like off-diagonal fields} respectively in the two cases.

\section{Simplified version of the ansatz}
\label{section_simple_ansatz}
In this section we give an ansatz for the $S^J$ matrix of a $(J,\psi)-$extended permutation orbifold in the case where the mother theory does not have fixed points and hence we can forget about orbifold diagonal fields \cite{Maio:2009kb}. Since fixed points arise only for (half-)integer spin simple currents, examples when this happens are all those mother theories which admit simple currents with $h_J\in\frac{1}{4}\mathbb{Z}_{\rm odd}$ and that generate half-integer spin currents in the permutation orbifold.

This is our ansatz for the $(J,\psi)$-extended $\mathbb{Z}_2$-permutation orbifold when the mother theory has no fixed points of the current $J$.
\begin{subequations}
\label{ansatz with no fixed points}
\begin{eqnarray}
S^{(J,\psi)}_{(mn)(pq)}&=&0 \\
S^{(J,\psi)}_{(mn)\widehat{(p,\chi)}}&=&A\cdot S_{mp} \\
S^{(J,\psi)}_{\widehat{(m,\phi)}\widehat{(p,\chi)}}&=&B\cdot\frac{1}{2}\,e^{i\pi \hat{Q}_J(m)}\,P_{J\cdot m,p}\, \,e^{i\pi(\phi+\chi)}
\end{eqnarray}
\end{subequations}
where $A$ and $B$ are phases that will be constrained by modular invariance to be equal to
\begin{equation}
A=i^\psi\,e^{i\pi\,h_J}\qquad\&\qquad B=(-1)^\psi\,e^{3i\pi\,h_J}
\end{equation}
We will prove this in the next subsection. Actually, the phase $A$ is determined up to a sign: here we have made the choice of taking the positive sign. These sign choices are a remnant of the sign ambiguities one has in general in matrices $S^J$. Most of
the relative sign choices are fixed within blocks of the matrix, because we write it in terms of $S$ and $P$, but one off-diagonal
choice between two blocks remains. 

The notation in the ansatz is as follows. We denote by $\hat{Q}_J (m)$ the combination of weights $\hat{Q}_J (m)= h_J+h_m-h_{J\cdot m}$, while $Q_J(m)$ is the monodromy charge of the field $m$ w.r.t. the current $J$ in the mother theory which gives rise to the current $(J,\psi)$ in the permutation orbifold (independently of its symmetric or anti-symmetric representation). These two quantities are obviously related by $Q_J(m)=\hat{Q}_J(m) \,\,{\rm mod}\,\, \mathbb{Z}$.

This ansatz more or less interpolates our previous results \cite{Maio:2009kb,Maio:2009cy}, up to some sign related to the ambiguities one has in choosing the order of the splitted fixed points. The phase in the twisted-twisted sector containing the hatted monodromy charge is necessary in order to make $S^J$ symmetric\footnote{In fact one can check that
\begin{equation}
e^{i\pi \hat{Q}_J(m)}\,P_{Jm,p}=e^{i\pi \hat{Q}_J(p)}\,P_{m,Jp} \equiv A_{mp}\nonumber
\end{equation}
with 
\begin{equation}
A_{mp}= e^{i\pi h_J}\,\sqrt{T}_{mm}\sum_l\left(e^{2 i\pi Q_J(l)}\,S_{ml}T^2_{ll}S_{lp}\right) \sqrt{T}_{pp}
\nonumber
\end{equation}
and $A_{mp}$ is symmetric.} as it should be for order-two currents, since in general $S^J_{ab}=S^{J^{-1}}_{ba}$ \cite{Fuchs:1996dd}. We need to put a hat on $Q_J$ in order to avoid ambiguities deriving from having the monodromy charge in the exponent, since it is defined only modulo integers. Similarly to what happens in the BHS formula \cite{Borisov:1997nc}, the $P$ matrix enters the twisted-twisted sector.

Observation: in \cite{Maio:2009kb} we derived an explicit formula for $S^{(J,\psi)\equiv (0,1)}_{(mn)(pq)}$, which was
\begin{equation}
S^{(J,\psi)\equiv (0,1)}_{(mn)(pq)}=S_{mp}\,S_{nq}-S_{mq}\,S_{np}\,.
\end{equation}
In the first line of our ansatz (\ref{ansatz with no fixed points}), instead, we have put $S^{(J,\psi)}_{(mn)(pq)}=0$. How can we combine these two apparently different formulas? Actually, this ansatz does not apply to the identity field ``$0$'' since it does have fixed points, namely all the fields of the mother theory. That might be the reason why the first line looks exceptional. However, it is very tempting to guess that the ansatz should be something like  $S^{(J,\psi)}_{(mn)(pq)}= S^J_{mp}\,S^J_{nq}-S^J_{mq}\,S^J_{np}$ for the general case when $J$ has fixed points in the mother theory, since it looks very much like our first expression and, moreover, it reduces to $S^{(J,\psi)}_{(mn)(pq)}=0$ when $J$ has no fixed points. We will see later that this is indeed the case, provided we make a $\psi-$dependent sign change in our guess. Of course that still leaves the diagonal fields to worry about.

In order to give a flavor to the reader of the genre of calculations we have to perform to check the correctness of our ansatz, in the next two subsections we prove unitarity and modular invariance of $S^{(J,\psi)}$. These calculation are easy in this simplified case and for this reason we will do it here. They become more lengthy in the general situation, where a few tricks are necessary, and we will do it in the appendix.

\subsection{Unitarity of the simplified ansatz}
In this subsection we prove that the ansatz (\ref{ansatz with no fixed points}) gives a unitary $S^{(J,\psi)}$ matrix. For this we need to compute three matrix elements. Since the current $J$ has no fixed points in the mother theory, the off-diagonal fields are only orbit-like. The calculation is pretty straightforward, up to a few aside results that we will refer to as \textit{Corollary 1} and \textit{Corollary 2}, proven in the appendix (see (\ref{Corollary 1 app}) and (\ref{Corollary 2 app})). We recall them here:\\
\noindent \underline{\textbf{Corollary 1}}.\\
\begin{equation}
\sum_{p {\,\,\rm s.t.\,\,} \widehat{(p,\chi)} {\rm \,\,f.p.\,\,of\,\,} (J,\psi)} 
S_{ip}\,S_{pj}^\star
=\frac{1}{2}\,\delta_{ij} +  (-1)^{\psi+4h_J} \frac{1}{2}\,\delta_{Ji,j}\,.
\end{equation}
\noindent \underline{\textbf{Corollary 2}}.\\
\begin{equation}
\label{Corollary 2}
\sum_{a {\,\,\rm s.t.\,\,} \widehat{(a,\chi)} {\rm \,\,f.p.\,\,of\,\,} (J,\psi)} 
P_{p,Ja}\,P_{Ja,q}^\star=\delta_{pq}
\end{equation}
(we have in mind the situation when $p$ and $q$ give rise to twisted fields in the extension of the permutation orbifold, i.e. $Q_J(q)=\frac{\psi}{2}+2h_J$).

Note that the sums are over a selection of fields and not over all the fields of the mother theory, hence we can solve it using a suitably defined \textit{projector} operator (see (\ref{projector PI})):
\begin{equation}
\Pi(p)=\frac{1}{2}\,\sum_{n=0}^{1}\,e^{2 i\pi n (Q_J(p)-\frac{\psi}{2}-2h_J)}\,.
\end{equation}
It is easy to see that
\begin{equation}
\Pi(p)=\left\{
\begin{array}{cl}
1 & {\,\,\rm if\,\,} Q_J(p)=\frac{\psi}{2}+2h_J\,\,\,({\rm mod}\,\,\mathbb{Z})\\
0 & {\,\,\rm if\,\,} Q_J(p)\neq\frac{\psi}{2}+2h_J\,\,\,({\rm mod}\,\,\mathbb{Z})\\
\end{array}
\right. 
\end{equation}

Now the strategy to show unitarity is to compute the quantity $S^{(J,\psi)}\cdot S^{(J,\psi)\dagger}$ and prove that it is equal to the unit matrix.\\
$i)$ Off-diagonal-Off-diagonal
\begin{eqnarray}
&& (S^{(J,\psi)}\cdot S^{(J,\psi)\dagger})_{(mn)(pq)}= \nonumber\\
&&= 
\sum_{(ab)} S^{(J,\psi)}_{(mn)(ab)}\,S^{(J,\psi)\dagger}_{(ab)(pq)}+
\sum_{\widehat{(a,\xi)}} S^{(J,\psi)}_{(mn)\widehat{(a,\xi)}}\,S^{(J,\psi)\dagger}_{\widehat{(a,\xi)}(pq)}=
\nonumber\\
&&=
0 +
\sum_{\xi=0}^1 \sum_{a,\,\widehat{(a,\xi)}\,\,{\rm f.p.of\,\,}(J,\psi)} 
S_{ma}\,S_{ap}^\star=
\nonumber\\
&&=
2 \sum_{a,\,\widehat{(a,\xi)}\,\,{\rm f.p.of\,\,}(J,\psi)} S_{ma}\,S^\star_{ap}=
2\left[\frac{1}{2}\,\delta_{mp}+(-1)^{\psi+4h_J}\frac{1}{2}\,\delta_{Jm,p}\right]=\delta_{mp}\,.\nonumber
\end{eqnarray}
We have used the fact that $\delta_{Jm,p}=0$, since $m<n$ and $p<q$. The term within squared brackets comes from Corollary 1.\\
$ii)$ Off-diagonal-Twisted
\begin{eqnarray}
&& (S^{(J,\psi)}\cdot S^{(J,\psi)\dagger})_{(mn)\widehat{(p,\chi)}}= \nonumber\\
&&=
\sum_{(ab)} S^{(J,\psi)}_{(mn)(ab)}\,S^{(J,\psi)\dagger}_{(ab)\widehat{(p,\chi)}}+
\sum_{\widehat{(a,\xi)}} S^{(J,\psi)}_{(mn)\widehat{(a,\xi)}}\,S^{(J,\psi)\dagger}_{\widehat{(a,\xi)}\widehat{(p,\chi)}}=\nonumber\\
&&=
0 + A\,B^\star\,\sum_{\xi=0}^1 \sum_{a,\,\widehat{(a,\xi)}\,\,{\rm f.p.of\,\,}(J,\psi)} 
S_{ma} \,\frac{1}{2}\,e^{-i\pi\hat{Q}_J(a)}\,P^\star_{Ja,p}\,e^{i\pi(\xi+\chi)}=
0\,.\nonumber
\end{eqnarray}
The last sum vanishes because of $\sum_\xi e^{i\pi\xi}=0$.\\
$iii)$ Twisted-Twisted
\begin{eqnarray}
&& (S^{(J,\psi)}\cdot S^{(J,\psi)\dagger})_{\widehat{(p,\chi)}\widehat{(q,\phi)}}= \nonumber\\
&&=
\sum_{(mn)} S^{(J,\psi)}_{\widehat{(p,\chi)}(mn)}\,S^{(J,\psi)\dagger}_{(mn)\widehat{(q,\phi)}}+
\sum_{\widehat{(a,\xi)}} S^{(J,\psi)}_{\widehat{(p,\chi)}\widehat{(a,\xi)}}\,S^{(J,\psi)\dagger}_{\widehat{(a,\xi)}\widehat{(q,\phi)}}=
\nonumber\\
&&=
\sum_{(mn)} S_{pm} \,S_{mq}^\star  +\nonumber\\
&&+
\sum_{\xi=0}^1 \sum_{a,\,\widehat{(a,\xi)}\,\,{\rm f.p.of\,\,}(J,\psi)} 
\frac{1}{2}\,e^{i\pi\hat{Q}_J(a)}\, P_{p,Ja}\,e^{i\pi(\chi+\xi)}\,\frac{1}{2}\,e^{-i\pi\hat{Q}_J(a)}\, P^\star_{Ja,q}\,e^{i\pi(\xi+\phi)}=\nonumber\\
&&=
\sum_{(mn), n=Jm} S_{pm}\,S^\star_{mq}+
\frac{1}{2}\,e^{i\pi(\chi+\phi)}\,\sum_{a,\,\widehat{(a,\xi)}\,\,{\rm f.p.of\,\,}(J,\psi)} P_{p,Ja}\,P^\star_{Ja,q}\nonumber=\\
&&=
\frac{1}{2}\sum_{m} S_{pm}\,S^\star_{mq}+
e^{i\pi(\chi+\phi)}\,\frac{1}{2}\,\delta_{pq}=
\frac{1}{2}\,\delta_{pq} + \frac{1}{2}\,e^{i\pi(\chi+\phi)}\,\delta_{pq}= \delta_{pq}\,\delta_{\chi\phi}\,.
\nonumber
\end{eqnarray}
In the last lines we have replaced $\sum_{(mn)}=\frac{1}{2}\,\sum_m$, exploited unitarity of the original $S_{mp}$ and used Corollary 2.

Together, $i)$, $ii)$ and $iii)$ say that $S^{(J,\psi)}$ is unitary.

\subsection{Modular invariance of the simplified ansatz}
In order to check modular invariance, we will use the relation $(S^{(J,\psi)}T^{(J,\psi)})^3=(S^{(J,\psi)})^2$, but rewritten as
\begin{equation}
{T^{(J,\psi)}}^{-1} S^{(J,\psi)} {T^{(J,\psi)}}^{-1}=S^{(J,\psi)} T^{(J,\psi)} S^{(J,\psi)}\,.
\end{equation}
This is a much more convenient expression since involves no sums on the l.h.s. and only one sum on the r.h.s. We will have to prove this identity sector by sector. It will be useful to recall here equation (\ref{P_(Jp,q)}) for the $P$ matrix with one $J-$translated index:
\begin{equation}
P_{Jp,q}=e^{i\pi(h_J-\hat{Q}_J(p))}\,\sqrt{T}_p\,
\sum_m e^{2 i\pi Q_J(m)}\,S_{pm}\,T^2_m\,S_{mq}\,\sqrt{T}_q\,,
\end{equation}
where $\hat{Q}_J(p)=h_J+h_p-h_{Jp}$. We will use the notation $T_{ij}=T_i\,\delta_{ij}$ for the (diagonal) $T$ matrix.\\
\\
$i)$ Off-diagonal-Off-diagonal\\
l.h.s.:
\begin{eqnarray}
&& \left({T^{(J,\psi)}}^{-1} S^{(J,\psi)} {T^{(J,\psi)}}^{-1}\right)_{(mn)(pq)}=
{T^{(J,\psi)}}^{-1}_{(mn)} S^{(J,\psi)}_{(mn)(pq)} {T^{(J,\psi)}}^{-1}_{(pq)}=
0\,.\nonumber
\end{eqnarray}
r.h.s.:
\begin{eqnarray}
&& \left(S^{(J,\psi)} T^{(J,\psi)} S^{(J,\psi)}\right)_{(mn)(pq)}=\nonumber\\
&&=
\sum_{(ab)} S^{(J,\psi)}_{(mn)(ab)} T^{(J,\psi)}_{(ab)} S^{(J,\psi)}_{(ab)(pq)}+ 
\sum_{\widehat{(a,\xi)}} S^{(J,\psi)}_{(mn)\widehat{(a,\xi)}} T^{(J,\psi)}_{\widehat{(a,\xi)}} S^{(J,\psi)}_{\widehat{(a,\xi)}(pq)}=\nonumber\\
&&=
0+
\sum_{\xi=0}^1 \sum_{a,\,\widehat{(a,\xi)}\,\,{\rm f.p.of\,\,}(J,\psi)} \,
A^2\,S_{ma} \,\sqrt{T}_a \,e^{i\pi\xi}\,S_{ap} =0\,.\nonumber
\end{eqnarray}
The last equality comes from $\sum_{\xi=0}^1 e^{i\pi\xi}\,=0$. Hence: 
\begin{equation}
\left(S^{(J,\psi)} T^{(J,\psi)} S^{(J,\psi)}\right)_{(mn)(pq)}=
\left({T^{(J,\psi)}}^{-1} S^{(J,\psi)} {T^{(J,\psi)}}^{-1}\right)_{(mn)(pq)}\,.
\end{equation}
\\
$ii)$ Off-diagonal-Twisted\\
l.h.s.:
\begin{eqnarray}
&& \left({T^{(J,\psi)}}^{-1} S^{(J,\psi)} {T^{(J,\psi)}}^{-1}\right)_{(mn)\widehat{(p,\chi)}}=
{T^{(J,\psi)}}^{-1}_{(mn)} S^{(J,\psi)}_{(mn)\widehat{(p,\chi)}} {T^{(J,\psi)}}^{-1}_{\widehat{(p,\chi)}}= \nonumber\\
&&=
T^{-1}_m\,T^{-1}_n A\,S_{mp}\,\sqrt{T}^{-1}_p\,e^{-i\pi\chi}\,.\nonumber
\end{eqnarray}
r.h.s.:
\begin{eqnarray}
&& \left(S^{(J,\psi)} T^{(J,\psi)} S^{(J,\psi)}\right)_{(mn)\widehat{(p,\chi)}}=\nonumber\\
&&=
\sum_{(ab)} S^{(J,\psi)}_{(mn)(ab)} T^{(J,\psi)}_{(ab)} S^{(J,\psi)}_{(ab)\widehat{(p,\chi)}}+ 
\sum_{\widehat{(a,\xi)}} S^{(J,\psi)}_{(mn)\widehat{(a,\xi)}} T^{(J,\psi)}_{\widehat{(a,\xi)}} S^{(J,\psi)}_{\widehat{(a,\xi)}\widehat{(p,\chi)}}=\nonumber\\
&&=
0+
\sum_{\xi=0}^1 \sum_{a,\,\widehat{(a,\xi)}\,\,{\rm f.p.of\,\,}(J,\psi)} A\,
S_{ma}\,\sqrt{T}_a\,e^{i\pi\xi}\,
B\,\frac{1}{2}\,e^{i\pi\hat{Q}_J(a)}\,P_{Ja,p}\,e^{i\pi(\xi+\chi)}=\nonumber\\
&&=
A\,B \sum_{a,\,\widehat{(a,\xi)}\,\,{\rm f.p.of\,\,}(J,\psi)}
S_{ma}\,\sqrt{T}_a\,e^{i\pi h_J}\,\sqrt{T}_a\,\sum_l e^{2 i\pi Q_J(l)}\, S_{al}\,T^2_l\,S_{lp}\,\sqrt{T}_p\,e^{i\pi\chi}=\nonumber\\
&&=
A\,B\,e^{i\pi h_J} \sum_l\left(
\sum_{a,\,\widehat{(a,\xi)}\,\,{\rm f.p.of\,\,}(J,\psi)}
S_{ma}\,T_a\,S_{al}\right) 
e^{2 i\pi Q_J(l)}\,T^2_l\,S_{lp}\,\sqrt{T}_p\,e^{i\pi\chi}\,.\nonumber
\end{eqnarray}
We used (\ref{P_(Jp,q)}) in the third line. The sum within round brackets in the last line can be further simplified by using the projector as in (\ref{projector PI}) and the relation $T^{-1}ST^{-1}=STS$ for the $S$ and $T$ matrices of the original theory: in fact,
\begin{eqnarray}
&& \sum_{a,\,\widehat{(a,\xi)}\,\,{\rm f.p.of\,\,}(J,\psi)}
S_{ma}\,T_a\,S_{al}=
\sum_a \Pi(a)\,
S_{ma}\,T_a\,S_{al}=\nonumber\\
&&=
\sum_a \frac{1}{2}\,\sum_{k=0}^1 e^{2 i\pi k(Q_J(a)-\frac{\psi}{2}-2h_J)}\,
S_{ma}\,T_a\,S_{al}=\nonumber\\
&&=
\frac{1}{2}\,\sum_{k=0}^1 e^{-2 i\pi k(\frac{\psi}{2}+2h_J)}\,
\sum_a S_{J^k m,a}\,T_a\,S_{al}=\nonumber\\
&&=
\frac{1}{2}\,\sum_a S_{ma}\,T_a\,S_{al}+\frac{1}{2}\,(-1)^{\psi+4h_J}\,\sum_a S_{na}\,T_a\,S_{al}=\nonumber\\
&&=
\frac{1}{2}\,\left[(STS)_{ml}+(-1)^{\psi+4h_J} (STS)_{nl}\right]=\nonumber\\
&&=
\frac{1}{2}\,\left[(T^{-1}ST^{-1})_{ml}+
(-1)^{\psi+4h_J} (T^{-1}ST^{-1})_{nl}\right]\,.\nonumber
\end{eqnarray}
Then, going back to our main expression and recalling that $S_{Ji,j}=e^{2i\pi Q_J(j)}\,S_{ij}$ \cite{Schellekens:1990xy}, we have:
\begin{eqnarray}
&& \left(S^{(J,\psi)} T^{(J,\psi)} S^{(J,\psi)}\right)_{(mn)\widehat{(p,\chi)}}=\dots=
\nonumber\\
&&=
A\,B\, e^{i\pi h_J}\,\frac{1}{2}\sum_l\,
[T^{-1}_m\,S_{ml}\,T^{-1}_l+(-1)^{\psi+4h_J}\,T^{-1}_n\,S_{nl}\,T^{-1}_l]\,
e^{2i\pi Q_J(l)}\,T^2_l\,S_{lp}\,\sqrt{T}_p\,e^{i\pi\chi}=\nonumber\\
&&=
A\,B\, e^{i\pi h_J}\,\frac{1}{2}\sum_l\,
[T^{-1}_m\,S_{Jm,l}\,T^{-1}_l+(-1)^{\psi+4h_J}\,T^{-1}_n\,S_{Jn,l}\,T^{-1}_l]\,
T^2_l\,S_{lp}\,\sqrt{T}_p\,e^{i\pi\chi}=\nonumber\\
&&=
A\,B\, e^{i\pi h_J}\,\frac{1}{2}\,
[T^{-1}_m\,(STS\sqrt{T})_{np}+(-1)^{\psi+4h_J}\,T^{-1}_n\,(STS\sqrt{T})_{mp}]\,e^{i\pi\chi}=\nonumber\\
&&=
A\,B\, e^{i\pi h_J}\,\frac{1}{2}\,
[T^{-1}_m\,T^{-1}_n\,S_{np}\,\sqrt{T}^{-1}_p+ (-1)^{\psi+4h_J}\,T^{-1}_n\,T^{-1}_m\,S_{mp}\,\sqrt{T}^{-1}_p]\,e^{i\pi\chi}=\nonumber\\
&&=
A\,B\, e^{i\pi h_J}\,\frac{1}{2}\,T^{-1}_m\,T^{-1}_n\,(S_{np}+ (-1)^{\psi+4h_J}\,S_{mp})\,\sqrt{T}^{-1}_p\,e^{i\pi\chi}=\nonumber\\
&&=
A\,B\, (-1)^{\psi+4h_J}\,e^{i\pi h_J}\,T^{-1}_m\,T^{-1}_n\,S_{mp}\,\sqrt{T}^{-1}_p\,e^{i\pi\chi}\,.\nonumber
\end{eqnarray}
We have continuously used the relation $T^{-1}ST^{-1}=STS$ here. Hence
\begin{equation}
\left(S^{(J,\psi)} T^{(J,\psi)} S^{(J,\psi)}\right)_{(mn)\widehat{(p,\chi)}}=
\left({T^{(J,\psi)}}^{-1} S^{(J,\psi)} {T^{(J,\psi)}}^{-1}\right)_{(mn)\widehat{(p,\chi)}}\,,
\end{equation}
provided
\begin{equation}
B=(-1)^{\psi+4h_J}\,e^{-i\pi h_J}=(-1)^{\psi}\,e^{3i\pi h_J}\,.
\end{equation}
\\
$iii)$ Twisted-Twisted\\
l.h.s.:
\begin{eqnarray}
&& \left({T^{(J,\psi)}}^{-1} S^{(J,\psi)} {T^{(J,\psi)}}^{-1}\right)_{\widehat{(p,\phi)}\widehat{(q,\chi)}}=
{T^{(J,\psi)}}^{-1}_{\widehat{(p,\phi)}} S^{(J,\psi)}_{\widehat{(p,\phi)}\widehat{(q,\chi)}} {T^{(J,\psi)}}^{-1}_{\widehat{(q,\chi)}}= \nonumber\\
&&=
\frac{1}{2}\, B\,e^{i\pi\hat{Q}_J(p)}\,\sqrt{T}^{-1}_p \,P_{Jp,q}\,\sqrt{T}^{-1}_q \,.\nonumber
\end{eqnarray}
r.h.s.:
\begin{eqnarray}
&& \left(S^{(J,\psi)} T^{(J,\psi)} S^{(J,\psi)}\right)_{\widehat{(p,\phi)}\widehat{(q,\chi)}}=\nonumber\\
&&=
\sum_{(ab)} S^{(J,\psi)}_{\widehat{(p,\phi)}(ab)} T^{(J,\psi)}_{(ab)} S^{(J,\psi)}_{(ab)\widehat{(q,\chi)}}+ 
\sum_{\widehat{(a,\xi)}} S^{(J,\psi)}_{\widehat{(p,\phi)}\widehat{(a,\xi)}} T^{(J,\psi)}_{\widehat{(a,\xi)}} S^{(J,\psi)}_{\widehat{(a,\xi)}\widehat{(q,\chi)}}=\nonumber\\
&&=
\sum_{(ab)}
A^2\, S_{pa}\,T_a\,T_b\,S_{aq}+\nonumber\\
&&+
\sum_{\xi=0}^1 \sum_{a,\,\widehat{(a,\xi)}\,\,{\rm f.p.of\,\,}(J,\psi)} B^2\,
\frac{1}{2}\,e^{i\pi\hat{Q}_J(a)}\, P_{p,Ja}\,e^{i\pi(\phi+\xi)}\,\sqrt{T}_a\,e^{i\pi\xi}\,
\frac{1}{2}\,e^{-i\pi\hat{Q}_J(a)}\,P_{Ja,q}\,e^{i\pi(\xi+\chi)}=\nonumber\\
&&=
\frac{1}{2}\,A^2\,\sum_a S_{pa}\,T_a\,T_{Ja}\,S_{aq}+0\,.\nonumber
\end{eqnarray}
The contribution from the twisted-twisted sector vanishes due to $\sum_{\xi=0}^1 e^{i\pi\xi}=0$. Moreover, we can re-express $T_{Ja}$ in terms of $T_a$ using $T_{Ja}=T_a\,e^{2i\pi (h_J-\hat{Q}_J(a))}$. Hence:
\begin{eqnarray}
&& \left(S^{(J,\psi)} T^{(J,\psi)} S^{(J,\psi)}\right)_{\widehat{(p,\phi)}\widehat{(q,\chi)}}=\dots=\nonumber\\
&&=
\frac{1}{2}\,A^2\,\sum_a S_{pa}\,T^2_a\,e^{2i\pi (h_J-\hat{Q}_J(a))}\,S_{aq}=\nonumber\\
&&=
\frac{1}{2}\,A^2\,e^{2i\pi h_J}\,\sum_a e^{2 i\pi Q_J(a)}\, S_{pa}\,T^2_a\,S_{aq}=\nonumber\\
&&=
\frac{1}{2}\,A^2\,e^{i\pi h_J}\,e^{i\pi \hat{Q}_J(p)}\,
\sqrt{T}^{-1}_p\,P_{Jp,q}\,\sqrt{T}^{-1}_q
\end{eqnarray}
We have also used the freedom to replace $-\hat{Q}_J(a)$ with $Q_J(a)$ in the phase exponent appearing in the sum over $a$, which is allowed since both these quantities are either integer of half-integer and differ only by integer numbers. In the last line we have used (\ref{P_(Jp,q)}) to rewrite the sum over $a$ in terms of the $P$ matrix. Then, by comparison we get:
\begin{equation}
\left(S^{(J,\psi)} T^{(J,\psi)} S^{(J,\psi)}\right)_{\widehat{(p,\phi)}\widehat{(q,\chi)}}=
\left({T^{(J,\psi)}}^{-1} S^{(J,\psi)} {T^{(J,\psi)}}^{-1}\right)_{\widehat{(p,\phi)}\widehat{(q,\chi)}}\,,
\end{equation}
provided
\begin{equation}
B=A^2\,e^{i\pi h_J}\qquad \Longrightarrow \qquad A^2=(-1)^\psi \,e^{2i\pi h_J}\,.
\end{equation}

Together, $i)$, $ii)$ and $iii)$ say that $S^{(J,\psi)}$ is modular invariant. In addition, this calculation \textit{fully} fixes the phase $B$, while $A$ is fixed \textit{up to a sign}. For future convenience, we recall their values here:
\begin{equation}
A^2=(-1)^\psi \,e^{2i\pi h_J} \qquad\&\qquad B=(-1)^\psi \,e^{3i\pi h_J}\,.
\end{equation}

\section{The general ansatz}
\label{section_ansatz}
Here we extend our ansatz to the most general case, including when the simple currents of the mother theory admit fixed points, giving rise to the diagonal sector in the permutation orbifold. The most general ansatz is:
\begin{subequations}
\label{ansatz with fixed points}
\begin{eqnarray}
S^{(J,\psi)}_{(mn)(pq)}&=&S^J_{mp}\,S^J_{nq}+(-1)^\psi S^J_{mq}\,S^J_{np} \\
S^{(J,\psi)}_{(mn)\widehat{(p,\chi)}}&=&
\left\{
\begin{array}{cl}
0 & {\,\,\rm if\,\,}   J\cdot m=m\\
A\,S_{mp} & {\,\,\rm if\,\,}   J\cdot m=n
\end{array}
\right. \\
S^{(J,\psi)}_{\widehat{(p,\phi)}\widehat{(q,\chi)}}&=&
B\,\frac{1}{2}\,e^{i\pi\hat{Q}_J(p)}\,P_{Jp,q}\,e^{i\pi(\phi+\chi)} \\
S^{(J,\psi)}_{(i,\phi)(j,\chi)}&=&\frac{1}{2}\,S^J_{ij}\,S^J_{ij} \\
S^{(J,\psi)}_{(i,\phi)(mn)}&=&S^J_{im}\,S^J_{in} \\
S^{(J,\psi)}_{(i,\phi)\widehat{(p,\chi)}}&=&C\,\frac{1}{2}\,e^{i\pi\phi} \,S_{ip}\,.
\end{eqnarray}
\end{subequations}
Using modular invariance, we show in the appendix that these phases satisfy the following relations:
\begin{equation}
B=(-1)^\psi\,e^{3i\pi h_J}\,\,,\qquad A^2=C^2=(-1)^\psi\,e^{2 i\pi h_J}\,,
\end{equation}
$h_J$ being the weight of the simple current, which might depend on the central charge, rank and level of the original CFT. These relations comes from modular invariance: so, we can see that $B$ is \textit{fully} fixed, while $A$ and $C$ are fixed \textit{up to a sign}. We could 
also have inserted a phase $E$ also in the matrix element $S^{(J,\psi)}_{(i,\phi)(mn)}$. Modular invariance would then constrain it to $E^2=1$, hence $E$ would have been just a sign. As before in the simplified ansatz, these sign ambiguities are completely
understood in terms of the general sign ambiguities of fixed point resolution matrices. Within the three blocks
(diagonal, off-diagonal, twisted) they are fixed because we write all matrix elements in terms of $S^J$, $S$ and $P$, but this still
leaves three relative signs between the blocks. These signs are fixed by requiring that the result should recover the
BHS matrix. The latter has no free signs, because it is defined by a character representation. This therefore defines a
convenient canonical choice for the signs. The special case of the BHS formula corresponds to 
$h_J=\psi=0$ for the identity, hence $B=1$, while $A$ and $C$ are just signs, that must be taken positive.
However, we emphasize that any other sign
choice for $A$, $C$ or $E$ is equally valid; it is analogous to a gauge choice. Note that some of the matrices presented in our earlier work
\cite{Maio:2009kb}\cite{Maio:2009cy} use different sign conventions.

A comment about the matrix element $S^{(J,\psi)}_{(mn)\widehat{(p,\chi)}}$ is in order. We can actually \textit{prove} that the quantity $S_{mp}$ vanishes when $J\cdot m=m$ and $\psi=1$ and use the second line of the ansatz also in this case. In fact, first of all, in order for a fixed point of $J$ to exist $h_J$ must be (half-)integer and as a consequence we can drop the $2h_J$ addend from the monodromy of $p$, $Q_{J}(p)=\frac{\psi}{2}$; secondly, using $S_{Jm,p}=e^{2 i\pi Q_J(p)}\,S_{mp}$ \cite{Schellekens:1990xy}, we have:
\begin{equation}
S_{mp}=S_{Jm,p}=e^{2 i\pi Q_J(p)}\,S_{mp}=e^{2 i\pi \frac{\psi}{2}}\,S_{mp}\,,
\end{equation}
implying that the non-identically-to-zero option of $S^{(J,\psi)}_{(mn)\widehat{(p,\chi)}}$ actually also vanishes when $J\cdot m=m$ and $\psi=1$. So in our ansatz we are claiming that $S^{(J,\psi)}_{(mn)\widehat{(p,\chi)}}$ vanishes also for $\psi=0$ when $(mn)$ is fixed-point-like. We also recall that for orbit-like off-diagonal fields there exists a similar relation between $S_{mp}$ and $S_{np}$:
\begin{equation}
S_{np}=S_{Jm,p}=e^{2 i\pi Q_J(p)}\,S_{mp}=e^{2 i\pi \frac{\psi}{2}}\,S_{mp}\,,
\end{equation}
but we cannot infer much from here. It is crucial in these manipulations that the field $p$ gives rise to a twisted field in the extended orbifold.

\subsection{Unitarity and modular invariance}
Unitarity and modular invariance of the ansatz are referred to the appendix. The calculation is cumbersome and some machinery needs to be developed. Nevertheless, we would like to stress a few facts about the calculation.

In order to prove unitarity, we show that 
\begin{equation}
S^{(J,\psi)}\cdot S^{(J,\psi)\dagger}=1\,.
\end{equation}
The calculation is lengthy, but interesting since we are able to derive a few non-trivial aside identities that are collected in two corollaries (already stated before) having to do with projected sums of selected elements of the unitary matrices $S$ and $P$ of the original theory. The result is non-trivial but in the end simple.

Modular invariance is the statement that $(S^{(J,\psi)})^2= (S^{(J,\psi)} \cdot T^{(J,\psi)})^3$, where $T^{(J,\psi)}$ is the $T$ matrix of the permutation orbifold restricted to the fixed points of $(J,\psi)$. Using this relation to prove modular invariance would be computationally heavy, due to the double sum arising in the cube. Instead we re-write the constraint as
\begin{equation}
{T^{(J,\psi)}}^{-1} S^{(J,\psi)} {T^{(J,\psi)}}^{-1}=S^{(J,\psi)} T^{(J,\psi)} S^{(J,\psi)}\,,
\end{equation}
which is simpler since it involves only one sum on the r.h.s. and no sums at all on the l.h.s. Surprisingly enough, we find that the phases in the ansatz do not depend explicitly on the central charge $c$ of the mother CFT (the central charge of the permutation orbifold is $\hat{c}=2c$). The reason for this is that the $T$ matrices of the orbifold theory re-arrange themselves into suitable functions of $T$ matrices of the original theory. Explicitly (recall $T$ is diagonal: $T_{ij}=T_i\,\delta_{ij}$):
\begin{equation}
T^{(J,\psi)}_{(mn)}=T_m\,T_n\,,\qquad
T^{(J,\psi)}_{(i,\phi)}=T^2_i\,,\qquad
T^{(J,\psi)}_{\widehat{(p,\chi)}}=e^{i\pi\chi}\,\sqrt{T}_p\,.
\end{equation}
hence the central charge gets always re-absorbed in $T$. The phases $A$, $B$ and $C$ will be constrained by this calculation to be equal to the expressions given earlier.

\subsection{Checks}

Although we have an explicit proof that our results satisfy the conditions of modular invariance, we do not have a general proof
that all other RCFT conditions are satisfied, although the simplicity and generality of the answer suggests that this is indeed
the right answer. The next issue one could check is the fusion rules of the extended CFT. Currents of order two that have
fixed points must have integer or half-integer spin. In the latter case there is no extension, but one may consider instead
the tensor product with an Ising model, extended with an integer spin product of currents. Indeed, also for integer spin currents
one can consider arbitrarily complicated tensor products and any integer spin product current. All of these should give sensible
fusion rules. We have built (\ref{ansatz with fixed points}) into the program {\tt kac} \cite{kac}, which computes fusion rules
for simple current extended WZW models and coset CFT's, and this gives us access to a huge number of explicit examples. We have checked many simple extensions,
and also combinations of permutation orbifolds. For example, denote by $X$ the permutation orbifold of $C(3)_2$. It has
85 primaries and 
four simple currents, the identity, the anti-symmetric component of the latter (which has spin 1) and two spin 3 currents $K$ and $L$ originating
from symmetric and anti-symmetric product of the simple current of $C(3)_2$. We can now tensor $X$ with itself, and extend the result
with $(K,K)$ or $(K,L)$ or $(L,L)$. This gives three distinct CFT's with 2578, 2284 and 2102 primaries respectively. Checking all their
fusion rules is very time-consuming, so we have just checked a large sample.
The fusion rules we have checked in these cases, and many others, are indeed correct. Note that our formalism allows us to consider
also the permutation orbifold of $X \times X$, and the simple current extensions thereof. For all these CFT's the fusion rules are
now explicitly available. Furthermore, for all these cases we can compute the boundary and crosscap coefficients as well as 
the annulus, Moebius and Klein bottle amplitudes using the formalism of \cite{Fuchs:2000cm} (generalizing earlier
works, such as  \cite{Cardy:1989ir,Pradisi:1996yd,Pradisi:1995pp}, and references cited in this paper). 

A general proof of fusion rule integrality is beyond the scope of this paper, but may be doable. The essential ingredient should be
the observation at the basis of BHS, relating permutation orbifolds to orientifolds. Indeed, this is evident from the
appearance  of the $P$ matrix in the twisted field matrix elements. The proof of integrality is therefore presumably similar to
the proof of (annulus+Moebius) integrality in the case of orientifolds, as was argued in \cite{Borisov:1997nc} for the identity current. 
In \cite{Huiszoon:2002uj} such a proof was given for the orientifold formalism of \cite{Fuchs:2000cm} for all possible simple current
extensions. It is very likely that this proof can be adapted to cover the case of extensions of permutation orbifolds.

\section{Conclusion}
In this paper we have addressed the  problem of fixed point resolution in (extensions of) permutation orbifolds or equivalently the problem of finding the $S^J$ matrices for those classes of theories.

The $S^J$ matrices appear in many places both in conformal field theory and string theory. They are important in their own right as a tool in conformal field theory, in connection with the problem of resolving the fixed points in simple currents extensions of a given CFT. Via formula (\ref{main formula for f.p. resolution}), they provide us with the $S$ matrix of the extended theory, which  in turn gives the fusion coefficients of the product of two representations via the Verlinde formula. 

In string theory (and in boundary CFT), they also appear in several contexts. An example is the connection between the $S^J$ matrices with branes and boundary coefficients \cite{Fuchs:2000cm}. The most important application in string theory is probably when projections are involved (e.g. GSO or SUSY constraints): in fact, al sorts of projections are powerfully realized by simple current extensions of a given theory, hence the knowledge of the $S^J$ matrices becomes relevant.

This particular kind of orbifold that we are considering here might be interesting for phenomenological applications, especially in view of Gepner models. Gepner models \cite{Gepner:1987vz,Gepner:1987qi} are built out of (extensions of) tensor products of $N=2$ minimal models. A minimal model is a product of parafermionic theories \cite{Fateev:1985mm} $SU(2)_k/U(1)$ and a $SO(2)$ factor. In case (at least) two of the factor groups are the same, we can then consider the permutation orbifold arising from this $\mathbb{Z}_2$ symmetry that switches the factors around and extensions thereof.

The results of this paper allow us to make extensions of permutation orbifolds. We propose an ansatz for the $S^J$ matrices valid in the general case of simple currents of order 2. We have also shown how to get back the BHS formula when we extend the permutation orbifold by the identity current $(J,\psi)=(0,0)$. 
This ansatz is unitary and modular invariant. Moreover, unlike our previous results \cite{Maio:2009kb} and \cite{Maio:2009cy},
it does not depend 
on any explicit details of the
particular CFT used in the mother theory, other than its modular properties.
It depends only on the weight $h_J$ of the current used in the extension (via phases) and on the matrices $S$ and $P$ (or, equivalently, $T$) of the mother theory. This implies that it can be used freely in any sequence of extensions and $Z_2$ permutations of CFT's, thus
leading to a huge set of possible applications.

There are still further generalizations possible: the extension of this result to higher order permutations and the extension to higher order
currents, and the combination of both. 

\section*{Acknowledgments}
This research is supported by the Dutch Foundation for Fundamental Research of Matter (FOM)
as part of the program STQG (String Theory and Quantum Gravity, FP 57). This work has been partially 
supported by funding of the Spanish Ministerio de Ciencia e Innovaci\'on, Research Project
FPA2008-02968, and by the Project CONSOLIDER-INGENIO 2010, Programme CPAN
(CSD2007-00042).

\newpage
\appendix

\section{Twisted fixed points}
In \cite{Maio:2009kb} we showed that twisted fixed points $\widehat{(p,\xi)}$ of the current $(J,\psi)$ have monodromy charge $Q_J(p)=\frac{\psi}{2}$ (mod $\mathbb{Z}$) when the simple current $J$ has (half-)integer spin $h_J$. In this appendix we will show that for simple currents with spin $h_J=\frac{1}{4}\,\mathbb{Z}_{\rm odd}$ twisted fixed points have monodromy charge $Q_J(p)=\frac{1-\psi}{2}$ (mod $\mathbb{Z}$).

The starting point is the constraint (3.29) of \cite{Maio:2009kb}, which reads:
\begin{equation}
\label{constraint from first paper}
e^{i\pi\psi}\sum_n \frac{S_{Jn}\,P_{pn}\,P^\dagger_{np'}}{S_{0n}}=\delta_{pp'}\,.
\end{equation}
On the l.h.s. we can use $S_{Jl,n}=e^{2i\pi Q_J(n)}\,S_{ln}$ (which applies also to the case $l=0$, i.e. the identity) and expand the $P$ matrix using $P=\sqrt{T}\,S\,T^2\,S\,\sqrt{T}$. So we can write:
\begin{eqnarray}
{\rm l.h.s.}&=&
e^{i\pi\psi}\sum_{n,l,m} e^{2i\pi Q_J(n)}\,
\sqrt{T}_p\,S_{pl}\,T^2_l\,S_{ln}\,\sqrt{T}_n\cdot
\sqrt{T}^\star_n\,S^\star_{nm}\,{T^\star}^2_m\,S^\star_{mp'}\,\sqrt{T}^\star_{p'}=\nonumber\\
&=&
e^{i\pi\psi}\sum_{l,m}\,
\sqrt{T}_p\,S_{pl}\,T^2_l\cdot\delta_{Jl,m}\cdot {T^\star}^2_m\,S^\star_{mp'}\,\sqrt{T}^\star_{p'}=\nonumber\\
&=&
e^{i\pi\psi}\sum_{l}\,
\sqrt{T}_p\,S_{pl}\,T^2_l\, {T^\star}^2_{Jl}\,S^\star_{Jl,p'}\,\sqrt{T}^\star_{p'}=\nonumber\\
&=&
e^{i\pi\psi}\sum_{l}\,
\sqrt{T}_p\,S_{pl}\,\left[e^{2\pi i (h_l-h_{Jl})}\right]^2\,S^\star_{Jl,p'}\,\sqrt{T}^\star_{p'}=\nonumber\\
&=&
e^{i\pi\psi}\sum_{l}\,
\sqrt{T}_p\,S_{pl}\,\left[e^{2\pi i (h_J-Q_J(l))}\right]^2\,S^\star_{Jl,p'}\,\sqrt{T}^\star_{p'}=\nonumber\\
&=&
e^{i\pi\psi}\left[e^{2\pi i\,h_J}\right]^2\,\sum_{l}\,
\sqrt{T}_p\,S_{pl}\,S^\star_{Jl,p'}\,\sqrt{T}^\star_{p'}=\nonumber\\
&=&
e^{i\pi\psi}\left[e^{2\pi i\,h_J}\right]^2\,\sum_{l}\, e^{-2\pi iQ_J(p')}
\sqrt{T}_p\,S_{pl}\,S^\star_{lp'}\,\sqrt{T}^\star_{p'}=\nonumber\\
&=&
e^{i\pi\psi}\left[e^{2\pi i\,h_J}\right]^2\, e^{-2\pi iQ_J(p')}
\sqrt{T}_p\cdot\delta_{pp'}\cdot\sqrt{T}^\star_{p'}=\nonumber\\
&=&
e^{i\pi\psi}\left[e^{2\pi i\,h_J}\right]^2\, e^{-2\pi iQ_J(p)}\,\delta_{pp'}\,.\nonumber
\end{eqnarray}
Here we have removed the phase involving $Q_J(p)$ within squared brackets, since it is always (half-)integer for order-2 simple currents, so its square is one, but we have retained $h_J$ since it gives a non-trivial phase when $h_J\in\frac{1}{4}\, \mathbb{Z}_{\rm odd}$, while it is negligible when $h_J$ is (half-)integer. Hence our constraint (\ref{constraint from first paper}) becomes:
\begin{equation}
\label{twisted fp's}
e^{2\pi iQ_J(p)}= e^{i\pi\psi}\left[e^{2\pi i\,h_J}\right]^2
\qquad\Leftrightarrow\qquad 
Q_J(p)= \frac{\psi}{2}+2\,h_J \,\,\,({\rm mod}\,\,\mathbb{Z})\,.
\end{equation}
This is equivalent to say that
\begin{equation}
Q_J(p)=\left\{
\begin{array}{cl}
\frac{\psi}{2} & {\rm for}\,\, h_J\in\mathbb{Z} \,\,{\rm or}\,\, h_J\in\mathbb{Z}+\frac{1}{2}\\
\frac{1-\psi}{2} & {\rm for}\,\, h_J\in\frac{1}{4}\,\mathbb{Z}_{\rm odd}
\end{array}
\right.
\end{equation}

\newpage
\section{Unitarity}
In this section we prove unitarity of the ansatz via the relation $S^{(J,\psi)}\cdot S^{(J,\psi)\dagger}=1$. Before doing this we will need a few preliminary results that will be useful in the proof.

\subsection{Useful corollaries}
At some point we will need to compute
\begin{equation}
\sum_{p {\,\,\rm s.t.\,\,} \widehat{(p,\chi)} {\rm \,\,f.p.\,\,of\,\,} (J,\psi)} 
S_{ip}\,S_{pj}^\star\,.
\end{equation}
We will mostly consider (half-)integer spin simple currents $J$, for which $\widehat{(p,\chi)}$ is a fixed point of $(J,\psi)$ when $Q_J(p)=\frac{\psi}{2}$. The generalization to the case where $h_J\in\frac{1}{4}\,\mathbb{Z}_{\rm odd}$ is achieved by the shift
\begin{equation}
\psi\rightarrow \psi+4h_J
\qquad\leftrightarrow\qquad
(-1)^\psi\rightarrow (-1)^\psi e^{4\pi i\,h_J}\nonumber\,,
\end{equation}
but the current $J$ has no fixed points in this situation.

\noindent \underline{\textbf{Lemma}}.\\
We restrict ourselves to order-two simple currents, for which the monodromy charge of every field is either integer or half-integer. Define the \emph{projector}:
\begin{equation}
\Pi(p)=\frac{1}{2}\,\sum_{n=0}^{1}\,e^{2 i\pi n (Q_J(p)-\frac{\psi}{2}-2h_J)}\,.
\end{equation}
It is easy to see that
\begin{equation}
\label{projector PI}
\Pi(p)=\left\{
\begin{array}{cl}
1 & {\,\,\rm if\,\,} Q_J(p)=\frac{\psi}{2}+2h_J\,\,\,({\rm mod} \,\,\mathbb{Z})\\
0 & {\,\,\rm if\,\,} Q_J(p)\neq\frac{\psi}{2}+2h_J\,\,\,({\rm mod} \,\,\mathbb{Z})\\
\end{array}
\right.
\end{equation}
Since there are only two possibilities for the monodromy charge of order-two simple currents, we have either $Q_J(p)=\frac{\psi}{2}+2h_J\,\,\,({\rm mod} \,\,\mathbb{Z})$ or $Q_J(p)=1-\frac{\psi}{2}-2h_J\,\,\,({\rm mod} \,\,\mathbb{Z})$. 

In order to simplify the notation, we will consider (half-)integer spin simple currents, which is equivalent to say that we will drop the additional contribution $2\,h_J$ in our calculations, restoring it when necessary.

\noindent \underline{\textbf{Corollary 1}}.\\
We assume now that either $i$ or $j$ (not necessarily both) is a fixed point. For definiteness we take $i$ to be a fixed point of $J$. This would give rise to the diagonal field $(i,\phi)$ in the permutation orbifold as fixed point of the current $(J,\psi\equiv 0)$. Compute:
\begin{eqnarray}
\sum_{p {\,\,\rm s.t.\,\,} \widehat{(p,\chi)} {\rm \,\,f.p.\,\,of\,\,} (J,\psi)} 
S_{ip}\,S_{pj}^\star &=&
\sum_p \Pi(p)\, S_{ip}\,S_{pj}^\star = \nonumber\\
&=&\sum_p \,\frac{1}{2}\,\sum_{n=0}^{1}\,e^{2 i\pi n (Q_J(p)-\frac{\psi}{2})} S_{ip}\,S_{pj}^\star = \nonumber\\
&=&\sum_p \,\frac{1}{2}\,\sum_{n=0}^{1}\, e^{-2 i\pi n \frac{\psi}{2}} \, 
\underbrace{S_{J^n\cdot i,p}}_{=S_{ip}} \,S_{pj}^\star = \nonumber\\
&=&\underbrace{\frac{1}{2}\,\sum_{n=0}^{1}\, e^{-2 i\pi n \frac{\psi}{2}}}_{=\frac{1}{2}\,(1+(-1)^\psi)} \cdot
\underbrace{\sum_p \,S_{ip}\,S_{pj}^\star}_{=\delta_{ij}} = \frac{1}{2}\,(1+(-1)^\psi)\,\delta_{ij}\,. \nonumber
\end{eqnarray}
So we have found
\begin{equation}
\label{Corollary 1 with f.p.}
\sum_{p {\,\,\rm s.t.\,\,} \widehat{(p,\chi)} {\rm \,\,f.p.\,\,of\,\,} (J,\psi)} 
S_{ip}\,S_{pj}^\star \,\,=\,\, \frac{1}{2}\,(1+(-1)^\psi)\,\delta_{ij}\,,
\end{equation}
when either $i$ or $j$ is fixed point of $J$. 

What happens when neither $i$ nor $j$ is a fixed point of $J$? Using a similar reasoning we have:
\begin{eqnarray}
\sum_{p {\,\,\rm s.t.\,\,} \widehat{(p,\chi)} {\rm \,\,f.p.\,\,of\,\,} (J,\psi)} 
S_{ip}\,S_{pj}^\star &=&
\sum_p \Pi(p)\, S_{ip}\,S_{pj}^\star = \nonumber\\
&=&\sum_p \,\frac{1}{2}\,\sum_{n=0}^{1}\,e^{2 i\pi n (Q_J(p)-\frac{\psi}{2})} S_{ip}\,S_{pj}^\star = \nonumber\\
&=&\frac{1}{2}\sum_p \, 
\left[ S_{ip} \,S_{pj}^\star + (-1)^\psi S_{Ji,p} \,S_{pj}^\star \right]=
\nonumber\\
&=&\frac{1}{2}\,\delta_{ij} +  (-1)^\psi \frac{1}{2}\,\delta_{Ji,j}\,.
\end{eqnarray}
The generalized formula, valid also when $h_J\in\frac{1}{4}\,\mathbb{Z}_{\rm odd}$ is
\begin{equation}
\label{Corollary 1 app}
\sum_{p {\,\,\rm s.t.\,\,} \widehat{(p,\chi)} {\rm \,\,f.p.\,\,of\,\,} (J,\psi)} 
S_{ip}\,S_{pj}^\star =
\frac{1}{2}\,[\delta_{ij} +  (-1)^\psi e^{4\pi i\,h_J} \,\delta_{Ji,j}]\,.
\end{equation}

Observe that in case either $i$ or $j$ is a fixed point\footnote{In order for $i$ to be a fixed point of $J$, $h_J$ must be (half-)integer.} of $J$, formula (\ref{Corollary 1 app}) reduces to (\ref{Corollary 1 with f.p.}). Diagonal fixed points in the extended permutation orbifold arise from fixed points of the original theory when $\psi=0$: in this case this formula gives a Dirac delta
\begin{equation}
\sum_{p {\,\,\rm s.t.\,\,} \widehat{(p,\chi)} {\rm \,\,f.p.\,\,of\,\,} (J,\psi)} 
S_{ip}\,S_{pj}^\star=\delta_{ij}\,,\qquad (Ji=i \,\,\&\,\, \psi=0) \,.
\nonumber
\end{equation}
On the contrary, when $\psi=1$, $\sum_{p {\,\,\rm s.t.\,\,} \widehat{(p,\chi)} {\rm \,\,f.p.\,\,of\,\,} (J,\psi)} S_{ip}\,S_{pj}^\star =0$. Actually, even more strongly, one can prove that each term in the sum vanishes separately:
\begin{equation}
\label{S=0 if Jm=m and psi=1}
S_{ip}=0 \qquad {\rm if} \qquad Ji=i \,\,\&\,\, \psi=1\,.
\end{equation}
The proof is in the following identity:
\begin{equation}
S_{ip}=S_{Ji,p}=e^{2 i\pi Q_J(p)}\,S_{ip}=e^{2 i\pi\frac{\psi}{2}}\,S_{ip}\,
\end{equation}
hence, for $\psi=1$, $S_{ip}=0$.

As consequence of this result, we have another relation that will also be useful:\\

\noindent \underline{\textbf{Corollary 2}}.\\
We want to compute:
\begin{equation}
\sum_{a {\,\,\rm s.t.\,\,} \widehat{(a,\chi)} {\rm \,\,f.p.\,\,of\,\,} (J,\psi)} 
P_{p,Ja}\,P_{Ja,q}^\star\,.
\end{equation}
It will be useful in particular to consider the case when $p$ and/or $q$ give rise to twisted fields in the extended orbifold.
For (half-)integer spin simple currents this is the same as computing
\begin{equation}
\sum_{a {\,\,\rm s.t.\,\,} \widehat{(a,\chi)} {\rm \,\,f.p.\,\,of\,\,} (J,\psi)} 
P_{pa}\,P_{aq}^\star\,,
\end{equation}
since if $\widehat{(a,\chi)}$ is a fixed point, then also $\widehat{(Ja,\chi)}$ is such, due to the monodromy charge conservation $Q_J(Ja)=Q_J(J)+Q_J(a)$.
Now, expand $P$ in terms of $S$ and $T$ (the sum over repeated indices is understood):
\begin{subequations}
\begin{eqnarray}
P_{p,Ja}&=&\sqrt{T}_{pp}\,S_{pm}\,T^2_{mm}\,S_{m,Ja}\,\sqrt{T}_{Ja,Ja}\\
P_{Ja,q}^\star&=&(\sqrt{T}_{Ja,Ja}\,S_{Ja,n}\,T^2_{nn}\,S_{nq}\,\sqrt{T}_{qq})^\star\,.
\end{eqnarray}
\end{subequations}
Then
\begin{eqnarray}
&&\sum_{a {\,\,\rm s.t.\,\,} \widehat{(a,\chi)} {\rm \,\,f.p.\,\,of\,\,} (J,\psi)} 
P_{p,Ja}\,P_{Ja,q}^\star= \nonumber\\
&&=
\sum_m \sum_n \sqrt{T}_{pp}S_{pm}T^2_{mm}\,
\left(\sum_{a {\,\,\rm s.t.\,\,} \widehat{(a,\chi)} {\rm \,\,f.p.\,\,of\,\,} (J,\psi)}
S_{m,Ja}\,S^\star_{Ja,n} \right)
T^{2\star}_{nn}\,S^\star_{nq}\sqrt{T}^\star_{qq} =\nonumber\\
&&=
\sum_m \sum_n \sqrt{T}_{pp}\, S_{pm}\,T^2_{mm}
\left(\frac{1}{2}\delta_{mn}+(-1)^\psi\frac{1}{2}\delta_{Jm,n}\right)
T^{2\star}_{nn}\,S^\star_{nq}\,\sqrt{T}^\star_{qq} \,.\nonumber
\end{eqnarray}
In going from the second line to the third line we have used the freedom to re-shuffle the fields in the sum $\sum_{a {\,\,\rm s.t.\,\,} \widehat{(a,\chi)} {\rm \,\,f.p.\,\,of\,\,} (J,\psi)} S_{m,Ja}\,S^\star_{Ja,n}$, which is then equal to the sum $\sum_{Ja {\,\,\rm s.t.\,\,} \widehat{(Ja,\chi)} {\rm \,\,f.p.\,\,of\,\,} (J,\psi)} S_{m,a}\,S^\star_{a,n}$ for (half-)integer spin currents. After a few simplifications, the first term gives a half Dirac delta, while the second term contains a factor $\sum_m S_{pm}\,S^\star_{Jm,q}= e^{-2i\pi Q_J(q)}\delta_{mq}$. The latter delta allows us to make some extra simplifications. Moreover, if $q$ gives rise to the twisted field $\widehat{(q,\chi)}$ in the $(J,\psi)-$extended permutation orbifold, the power $e^{-2i\pi Q_J(q)}$ cancels the phase $(-1)^\psi$. In this latter case, we have
\begin{equation}
\label{Corollary 2 app}
\sum_{a {\,\,\rm s.t.\,\,} \widehat{(a,\chi)} {\rm \,\,f.p.\,\,of\,\,} (J,\psi)} 
P_{p,Ja}\,P_{Ja,q}^\star= 
\frac{1}{2}\,\delta_{pq} + \frac{1}{2}\,\delta_{pq} \,T^2_m \, T^{2\star}_{Jm}=\delta_{pq}\,,
\end{equation}
with $T^2_m \, T^{2\star}_{Jm}=[e^{2i\pi (h_m-h_{Jm})}]^2=[e^{2i\pi (Q_J(m)-h_J)}]^2=1$, coming from the fact that $h_J$ is (half-)integer and $Q_J(m)$ is also (half-)integer for order-two simple currents.

For simple currents with $h_J\in\frac{1}{4}\,\mathbb{Z}_{\rm odd}$ this reasoning is not valid anymore, since it is no longer true that if $\widehat{(a,\chi)}$ is a fixed point of $(J,\psi)$ then also $\widehat{(Ja,\chi)}$ is such. In this situation we need to do the calculation again, but the final answer will be still the same. In fact, in general we have
\begin{eqnarray}
&&\sum_{a {\,\,\rm s.t.\,\,} \widehat{(a,\chi)} {\rm \,\,f.p.\,\,of\,\,} (J,\psi)} S_{m,Ja}\,S^\star_{Ja,n}=\nonumber\\
&=&\left(\sum_{a {\,\,\rm s.t.\,\,} \widehat{(a,\chi)} {\rm \,\,f.p.\,\,of\,\,} (J,\psi)} S_{ma}\,S^\star_{an}\right)\,
e^{2\pi i Q_J(m)}\,e^{-2\pi i Q_J(n)}=\nonumber\\
&=&
\left(\frac{1}{2}\delta_{mn}+(-1)^{\psi+4h_J}\frac{1}{2}\delta_{Jm,n}\right)\,
e^{2\pi i Q_J(m)}\,e^{-2\pi i Q_J(n)}=\nonumber\\
&=&
\frac{1}{2}\,\delta_{mn}+(-1)^{\psi+4h_J}\frac{1}{2}\,e^{-4\pi i h_J}\,\delta_{Jm,n}=\nonumber\\
&=&
\frac{1}{2}\,\delta_{mn}+(-1)^{\psi}\frac{1}{2}\,\delta_{Jm,n}\,,\nonumber
\end{eqnarray}
where we have used $S_{m,Ja}=e^{2\pi i Q_J(m)}\,S_{ma}$; moreover, the phase come from the fact that $Q_J(Jm)=Q_J(J)+Q_J(m)$ and $Q_J(J)=2h_J$. Hence:
\begin{eqnarray}
&&\sum_{a {\,\,\rm s.t.\,\,} \widehat{(a,\chi)} {\rm \,\,f.p.\,\,of\,\,} (J,\psi)} 
P_{p,Ja}\,P_{Ja,q}^\star= \nonumber\\
&&=
\sum_m \sum_n \sqrt{T}_{p}S_{pm}T^2_{m}\,
\left(\sum_{a {\,\,\rm s.t.\,\,} \widehat{(a,\chi)} {\rm \,\,f.p.\,\,of\,\,} (J,\psi)}
S_{m,Ja}\,S^\star_{Ja,n} \right)
T^{2\star}_{n}\,S^\star_{nq}\sqrt{T}^\star_{q} =\nonumber\\
&&=
\sum_m \sum_n \sqrt{T}_{p}\, S_{pm}\,T^2_{m}
\left(\frac{1}{2}\delta_{mn}+(-1)^\psi\frac{1}{2}\delta_{Jm,n}\right)
T^{2\star}_{n}\,S^\star_{nq}\,\sqrt{T}^\star_{q} =\nonumber\\
&&=
\frac{1}{2}\delta_{pq}+\frac{1}{2}(-1)^\psi\sum_m
\sqrt{T}_{p}\, S_{pm}\,T^2_{m} 
T^{2\star}_{Jm}\,S^\star_{Jm,q}\,\sqrt{T}^\star_{q} =\nonumber\\
&&=
\frac{1}{2}\delta_{pq}+\frac{1}{2}(-1)^\psi e^{-4\pi i h_J} \sum_m
e^{-2\pi i Q_J(q)}\sqrt{T}_{p}\, S_{pm}\,S^\star_{mq}\,\sqrt{T}^\star_{q} =\nonumber\\
&&=
\frac{1}{2}\delta_{pq}+\frac{1}{2}(-1)^\psi e^{-4\pi i h_J}
e^{-2\pi i Q_J(q)}\,\delta_{pq} =\,\delta_{pq}\,,\nonumber
\end{eqnarray}
where we have used the fact that $Q_J(q)=\frac{\psi}{2}+2h_J$ and $(T_mT^\star_{Jm})^2= e^{-4\pi i h_J}$.

Both these corollaries will be useful in checking unitarity and this we will do now.

\subsection{Proof}
The strategy here to prove unitarity of the ansatz (\ref{ansatz with fixed points}) is to look at each sector separately and show that $S^{(J,\psi)}\cdot S^{(J,\psi)\dagger}=1$.

\subsubsection*{Diagonal-diagonal}
First of all, we have to recall that diagonal fields exists only for the symmetric representation of the current, namely $(J,\psi\equiv 0)$ and they come from fixed points $i$ in the mother theory, $J\cdot i=i$.

Now compute:
\begin{eqnarray}
&& (S^{(J,\psi)}\cdot S^{(J,\psi)\dagger})_{(i,\phi)(j,\chi)}= \nonumber\\
&&=
\sum_{(a,\xi)} S^{(J,\psi)}_{(i,\phi)(a,\xi)}\,S^{(J,\psi)\dagger}_{(a,\xi)(j,\chi)}+ 
\sum_{(mn)} S^{(J,\psi)}_{(i,\phi)(mn)}\,S^{(J,\psi)\dagger}_{(mn)(j,\chi)}+
\sum_{\widehat{(p,\xi)}}  S^{(J,\psi)}_{(i,\phi)\widehat{(p,\xi)}}\,S^{(J,\psi)\dagger}_{\widehat{(p,\xi)}(j,\chi)}=\nonumber\\
&&=
\sum_{\xi=0}^1 \sum_{a,\,Ja=a} \frac{1}{2}\,(S^J_{ia})^2\, \frac{1}{2}\,(S^{J\star}_{aj})^2+\nonumber\\
&&+
\sum_{(mn),m<n} S^J_{im}\,S^J_{in}\,S^{J\star}_{jm}\,S^{J\star}_{jn}+\nonumber\\
&&+
\sum_{\xi=0}^1 \sum_{p,\,\widehat{(p,\xi)}\,\,{\rm f.p.of\,\,}(J,\psi)} \frac{1}{2}\,S_{ip}\,e^{i\pi\phi}\, \frac{1}{2}\,S^{\star}_{pj}\,e^{-i\pi\chi}=\nonumber\\
&&=
\frac{1}{2}\sum_{a,Ja=a} (S^J_{ia})^2(S^{J\star}_{aj})^2+
\sum_{m,Jm=m}\sum_{n,Jn=n,n>m} S^J_{im}\,S^J_{in}\,S^{J\star}_{jm}\,S^{J\star}_{jn} +\frac{1}{2}\,\delta_{ij}\,e^{i\pi(\phi+\chi)}\,.\nonumber
\nonumber
\end{eqnarray}
In the last equality we have used the fact that $S^J_{ab}$ is non-zero only when both $a$ and $b$ are fixed points of $J$. The Kronecker delta comes from the third piece after using our previous Corollary 1 in the form of eq. (\ref{Corollary 1 with f.p.}). We have stressed the fact that in the off-diagonal fields $(m,n)$ $m<n$. This is important since the first and second piece can add up to give $\frac{1}{2}\,\delta_{ij}$. In fact, consider
\begin{eqnarray}
\frac{1}{2}\,\delta_{ij} &\equiv& 
\frac{1}{2}\left[\sum_{m}S^J_{im}S^{J\star}_{mj}\right]^2=
\frac{1}{2}\sum_{m,n} S^J_{im}S^{J\star}_{mj}S^J_{in}S^{J\star}_{nj}=\nonumber\\
&=&
\frac{1}{2}\left(\sum_{m<n}+\sum_{m=n}+\sum_{m>n}\right) S^J_{im}S^{J\star}_{mj}S^J_{in}S^{J\star}_{nj}=\nonumber\\
&=&
\frac{1}{2}\left(2\sum_{m<n}+\sum_{m=n}\right) S^J_{im}S^{J\star}_{mj}S^J_{in}S^{J\star}_{nj}=\nonumber\\
&=& \sum_{m<n} S^J_{im}S^{J\star}_{mj}S^J_{in}S^{J\star}_{nj} +\frac{1}{2} \sum_{a} S^J_{ia}S^{J\star}_{aj}S^J_{ia}S^{J\star}_{aj}\,,
\nonumber
\end{eqnarray}
which is exactly the term appearing in the first two contributions above. Hence
\begin{equation}
(S^{(J,\psi)}\cdot S^{(J,\psi)\dagger})_{(i,\phi)(j,\chi)}=
\frac{1}{2}\,\delta_{ij} \,+\frac{1}{2}\,e^{i\pi(\phi+\chi)}\,\delta_{ij}=\delta_{ij}\delta_{\phi\chi}\,,
\nonumber
\end{equation}
as it should be in order for $S^{(J,\psi)}$ to be unitary.

\subsubsection*{Diagonal-off-diagonal}
Again we consider here $\psi=0$ in order for diagonal fields to exist.

Compute:
\begin{eqnarray}
&& (S^{(J,\psi)}\cdot S^{(J,\psi)\dagger})_{(i,\phi)(mn)}= \nonumber\\
&&=
\sum_{(j,\xi)} S^{(J,\psi)}_{(i,\phi)(j,\xi)}\,S^{(J,\psi)\dagger}_{(j,\xi)(mn)}+ 
\sum_{(ab)} S^{(J,\psi)}_{(i,\phi)(ab)}\,S^{(J,\psi)\dagger}_{(ab)(mn)}+
\sum_{\widehat{(p,\xi)}}  S^{(J,\psi)}_{(i,\phi)\widehat{(p,\xi)}}\,S^{(J,\psi)\dagger}_{\widehat{(p,\xi)}(mn)}=\nonumber\\
&&=
\sum_{\xi=0}^1 \sum_{j,\,Jj=j} \frac{1}{2}\,(S^J_{ij})^2\, S^{J\star}_{jm}\,S^{J\star}_{jn}+\nonumber\\
&&+
\sum_{(ab)} S^J_{ia}\,S^J_{ib}\,\left( 
S^{J}_{am}\,S^{J}_{bn}+S^{J}_{an}\,S^{J}_{bm}\right)^\star+\nonumber\\
&&+
\sum_{\xi=0}^1 \sum_{p,\,\widehat{(p,\xi)}\,\,{\rm f.p.of\,\,}(J,\psi)} 
C\,\frac{1}{2}\,S_{ip}\,e^{i\pi\phi}\,\cdot
\left\{
\begin{array}{cl}
0 & {\,\,\rm if\,\,} Jm=m\\
A^\star\,S_{pm}^\star & {\,\,\rm if\,\,} Jm=n
\end{array}
\right. \nonumber
\end{eqnarray}
Now we have to distinguish two situations: \\
$i)$ $(mn)$ orbit-like, with $n=Jm$; \\
$ii)$ $(mn)$ fixed-point-like, with $m$ and $n$ fixed points of $J$.\\
We will see that in both cases the answer is zero, as it should be by unitarity.

$i)$ The first two lines give zero, since $S^J$ vanishes with one index equal to $m$ or $n$; from the third line we get
\begin{equation}
(S^{(J,\psi)}\cdot S^{(J,\psi)\dagger})_{(i,\phi)(mn)}=
C\,A^\star\,e^{i\pi\phi}
\underbrace{
\sum_{p,\,\widehat{(p,\xi)}\,\,{\rm f.p.of\,\,}(J,\psi)} S_{ip}\,S_{mp}^\star}_{=\delta_{im}\,\,
{\rm by\,\,Corollary\,\, 1}}=0\,,
\nonumber
\end{equation}
since the field $i\equiv Ji$ can never be equal to the field $m\neq Jm$.

$ii)$ The third line is now zero by the ansatz, while the other two give:
\begin{eqnarray}
&& (S^{(J,\psi)}\cdot S^{(J,\psi)\dagger})_{(i,\phi)(mn)}= \nonumber\\
&&=
\sum_{j,\,Jj=j} (S^J_{ij})^2\, S^{J\star}_{jm}\,S^{J\star}_{jn}+
\sum_{(ab),a<b} S^J_{ia}\,S^J_{ib}\,
\left(S^{J}_{am}\,S^{J}_{bn}+S^{J}_{an}\,S^{J}_{bm}\right)^\star=\nonumber\\
&&=
\sum_{j,\,Jj=j} (S^J_{ij})^2\, S^{J\star}_{jm}\,S^{J\star}_{jn}+
\sum_{a\neq b} S^J_{ia}\,S^J_{ib}\, S^{J\star}_{am}\,S^{J\star}_{bn}=\nonumber
\end{eqnarray}
Now consider the following equalities:
\begin{eqnarray}
\delta_{im}\delta_{in}&\equiv&
\sum_j \sum_l S^J_{ij}\,S^{J\star}_{jm}\, S^J_{il}\,S^{J\star}_{lm}=
\left(\sum_{j=l}+\sum_{j\neq l}\right) S^J_{ij}\,S^{J\star}_{jm}\, S^J_{il}\,S^{J\star}_{lm}=
\nonumber\\
&&=
\sum_j S^J_{ij}\,S^{J\star}_{jm}\, S^J_{ij}\,S^{J\star}_{jm}+
\sum_{j\neq l} S^J_{ij}\,S^{J\star}_{jm}\, S^J_{il}\,S^{J\star}_{lm}\,.\nonumber
\end{eqnarray}
Hence we can rewrite:
\begin{eqnarray}
&& (S^{(J,\psi)}\cdot S^{(J,\psi)\dagger})_{(i,\phi)(mn)}= \nonumber\\
&&=
\left(\delta_{im}\delta_{in}-
\sum_{j\neq l} S^J_{ij}\,S^{J\star}_{jm}\,S^J_{il}\,S^{J\star}_{ln}\right)+
\left(\sum_{a\neq b} S^J_{ia}\,S^J_{ib}\,S^{J\star}_{am}\,S^{J\star}_{bn}\right)=\nonumber\\
&&=
\,\delta_{im}\delta_{in}=0\,,
\nonumber
\end{eqnarray}
since $m\neq n$.

\subsubsection*{Diagonal-twisted}
Compute:
\begin{eqnarray}
&& (S^{(J,\psi)}\cdot S^{(J,\psi)\dagger})_{(i,\phi)\widehat{(p,\chi)}}= \nonumber\\
&&=
\sum_{(j,\xi)} S^{(J,\psi)}_{(i,\phi)(j,\xi)}\,S^{(J,\psi)\dagger}_{(j,\xi)\widehat{(p,\chi)}}+ 
\sum_{(mn)} S^{(J,\psi)}_{(i,\phi)(mn)}\,S^{(J,\psi)\dagger}_{(mn)\widehat{(p,\chi)}}+
\sum_{\widehat{(q,\xi)}}  S^{(J,\psi)}_{(i,\phi)\widehat{(q,\xi)}}\,S^{(J,\psi)\dagger}_{\widehat{(q,\xi)}\widehat{(p,\chi)}}=\nonumber\\
&&=
\sum_{\xi=0}^1 \sum_{j,\,Jj=j} \frac{1}{2}\,(S^J_{ij})^2\,
C^\star\,\frac{1}{2}\, S^{J\star}_{jp}\,e^{-i\pi\xi}+\nonumber\\
&&+
\sum_{(mn)} S^J_{im}\,S^J_{in}\, \cdot
\left\{
\begin{array}{cl}
0 & {\,\,\rm if\,\,} Jm=m\\
A^\star\,S_{mp}^\star & {\,\,\rm if\,\,} Jm=n
\end{array}
\right. +\nonumber\\
&&+
\sum_{\xi=0}^1 \sum_{q,\,\widehat{(q,\xi)}\,\,{\rm f.p.of\,\,}(J,\psi)} 
C\,\frac{1}{2}\,S_{iq}\,e^{i\pi\phi}\,B^\star\,\frac{1}{2}\,e^{-i\pi\hat{Q}_J(m)}\,P_{Jq,p}\, e^{i\pi(\xi+\chi)}=\nonumber\\
&&=
0\,+\,0\,+\,0\,=\,0\,,\nonumber
\end{eqnarray}
as it should be for unitarity. The first and third lines vanish because $\sum_{\xi=0}^1 e^{i\pi\xi}=0$. In the second line, the sum over all off-diagonal fields must be replaced by the sum over those diagonal fields with $m$ and $n$ fixed points, since otherwise $S^J$ vanishes; hence from the ansatz $S^{(J,\psi)\dagger}_{(mn)\widehat{(p,\chi)}}=0$.

\subsubsection*{Off-diagonal-off diagonal}
Compute:
\begin{eqnarray}
&& (S^{(J,\psi)}\cdot S^{(J,\psi)\dagger})_{(mn)(pq)}= \nonumber\\
&&=
\sum_{(i,\xi)} S^{(J,\psi)}_{(mn)(i,\xi)}\,S^{(J,\psi)\dagger}_{(i,\xi)(pq)}+ 
\sum_{(ab)} S^{(J,\psi)}_{(mn)(ab)}\,S^{(J,\psi)\dagger}_{(ab)(pq)}+
\sum_{\widehat{(a,\xi)}} S^{(J,\psi)}_{(mn)\widehat{(a,\xi)}}\,S^{(J,\psi)\dagger}_{\widehat{(a,\xi)}(pq)}=
\nonumber\\
&&=
\sum_{\xi=0}^1 \sum_{i,\,Ji=i} \,S^J_{mi}\,S^J_{ni}\,
S^{J\star}_{ip}\,S^{J\star}_{iq}+\nonumber\\
&&+
\sum_{(ab)} (S^J_{ma}\,S^J_{nb}+(-1)^\psi S^J_{mb}\,S^J_{na})
(S^J_{ap}\,S^J_{bq}+(-1)^\psi S^J_{aq}\,S^J_{bp})^\star +\nonumber\\
&&+
\sum_{\xi=0}^1 \sum_{a,\,\widehat{(a,\xi)}\,\,{\rm f.p.of\,\,}(J,\psi)}
\left\{
\begin{array}{cl}
0 & {\,\,\rm if\,\,} Jm=m \\
A\,S_{ma} & {\,\,\rm if\,\,} Jm=n
\end{array}
\right.
\cdot
\left\{
\begin{array}{cl}
0 & {\,\,\rm if\,\,} Jp=p \\
A^\star\,S^\star_{ap} & {\,\,\rm if\,\,} Jp=q
\end{array}
\right. \nonumber
\end{eqnarray}
We have to consider three cases:\\
$i)$ $(mn)$ and $(pq)$ are orbit-like;\\
$ii)$ $(mn)$ is orbit-like, $(pq)$ fixed-point-like;\\
$iii)$ $(mn)$ and $(pq)$ are fixed-point-like.\\
$i)$ 
\begin{eqnarray}
&&(S^{(J,\psi)}\cdot S^{(J,\psi)\dagger})_{(mn)(pq)}= \nonumber\\
&&=
0+0+2 \sum_{a,\,\widehat{(a,\xi)}\,\,{\rm f.p.of\,\,}(J,\psi)} S_{ma}\,S^\star_{ap}=
2\,\left(\frac{1}{2}\delta_{mp} + (-1)^{\psi+4h_J}\frac{1}{2}\delta_{Jm,p}\right)=\nonumber\\
&&=\delta_{mp}\,,\nonumber
\end{eqnarray}
consistently with unitarity. The first two contributions vanish, since $S^J$ vanishes, in the third term we have used Corollary 1. The $\psi-$dependent piece within brackets does not contribute since $\delta_{Jm,p}=0$, being $m<n$ and $p<q$.\\
$ii)$
\begin{eqnarray}
&&(S^{(J,\psi)}\cdot S^{(J,\psi)\dagger})_{(mn)(pq)}= \nonumber\\
&&=
0+0+0=0\nonumber\,.
\end{eqnarray}
This is in agreement with unitarity. The first two zeroes come from the $S^J=0$ for non-fixed points, the third from the ansatz.\\
$iii)$
\begin{eqnarray}
&&(S^{(J,\psi)}\cdot S^{(J,\psi)\dagger})_{(mn)(pq)}= \nonumber\\
&&=
2\,\delta_{\psi,0}\,\sum_{i,Ji=i} S^J_{mi}\,S^J_{ni}\,S^{J\star}_{ip}\,S^{J\star}_{iq}+\nonumber\\
&&+
\sum_{(ab),a<b}\left(S^J_{ma}\,S^J_{nb}+(-1)^\psi S^J_{mb}\,S^J_{na}\right)\,
\left(S^J_{ap}\,S^J_{bq}+(-1)^\psi S^J_{aq}\,S^J_{bp}\right)^\star+0\nonumber\,.
\end{eqnarray}
Recall that the first contribution is actually present only for $\psi=0$. The last contribution vanishes by the ansatz. The rest organizes in such a way to produce a delta:
\begin{eqnarray}
&&(S^{(J,\psi)}\cdot S^{(J,\psi)\dagger})_{(mn)(pq)}= \nonumber\\
&&=
2\,\delta_{\psi,0}\,\sum_{i,Ji=i} S^J_{mi}\,S^J_{ni}\,S^{J\star}_{ip}\,S^{J\star}_{iq}+\nonumber\\
&&+
\sum_{a\neq b} S^J_{ma}\,S^J_{nb}\,S^{J\star}_{ap}\,S^{J\star}_{bq}+
(-1)^\psi \sum_{a\neq b} S^J_{ma}\,S^J_{nb}\,S^{J\star}_{aq}\,S^{J\star}_{bp}\,.\nonumber
\end{eqnarray}
In fact, consider the following equalities:
\begin{eqnarray}
\delta_{mp}\,\delta_{nq} &\equiv& 
\sum_{i}\sum_{j} S^J_{mi}S^{J\star}_{ip}\,S^J_{nj}S^{J\star}_{jq}=\nonumber\\
&=&
\sum_{i} S^J_{mi}\,S^{J\star}_{ip}\,S^J_{ni}\,S^{J\star}_{iq}+
\sum_{i\neq j} S^J_{mi}\,S^{J\star}_{ip}\,S^J_{nj}\,S^{J\star}_{jq}\,.\nonumber
\end{eqnarray}
We can insert this into our expression and get
\begin{eqnarray}
&&(S^{(J,\psi)}\cdot S^{(J,\psi)\dagger})_{(mn)(pq)}= \nonumber\\
&&=
2\,\delta_{\psi,0}\,\left(\delta_{mp}\,\delta_{nq}-\sum_{i\neq j} S^J_{mi}\,S^{J\star}_{ip}\,S^J_{nj}\,S^{J\star}_{jq}\right)+\nonumber\\
&&+
\sum_{a\neq b} S^J_{ma}\,S^{J\star}_{ap}\,S^J_{nb}\,S^{J\star}_{bq}+
(-1)^{\psi}\sum_{a\neq b} S^J_{ma}\,S^{J\star}_{aq}\,S^J_{nb}\,S^{J\star}_{bp}\,.\nonumber
\end{eqnarray}
It is useful to consider the two cases $\psi=0$ and $\psi=1$ separately. Let's start with $\psi=0$. Then
\begin{eqnarray}
&&(S^{(J,\psi)}\cdot S^{(J,\psi)\dagger})_{(mn)(pq)}= \nonumber\\
&&=
2\,\delta_{mp}\,\delta_{nq}+\sum_{a\neq b} 
\left(
S^J_{ma}\,S^{J\star}_{aq}\,S^J_{nb}\,S^{J\star}_{bp}-
S^J_{ma}\,S^{J\star}_{ap}\,S^J_{nb}\,S^{J\star}_{bq}
\right)\,.
\nonumber
\end{eqnarray}
Here we can rewrite the sum as $\sum_{a\neq b}=\sum_{a,b}-\sum_{a=b}$: the first contribution is a product of two Dirac deltas while the second cancels out in the difference of the summands. Hence we have
\begin{equation}
(S^{(J,\psi)}\cdot S^{(J,\psi)\dagger})_{(mn)(pq)}=
2\,\delta_{mp}\,\delta_{nq}+\delta_{mq}\,\delta_{np}-\delta_{mp}\,\delta_{nq}=\delta_{mp}\,\delta_{nq}\,,
\nonumber
\end{equation}
since the other delta product vanishes, being $m<n$ and $p<q$. This agrees with unitarity.

Let's do the same calculation for $\psi=1$.
\begin{eqnarray}
&&(S^{(J,\psi)}\cdot S^{(J,\psi)\dagger})_{(mn)(pq)}= \nonumber\\
&&=
\sum_{a\neq b}
\left(
S^J_{ma}\,S^{J\star}_{ap}\,S^J_{nb}\,S^{J\star}_{bq}-
S^J_{ma}\,S^{J\star}_{aq}\,S^J_{nb}\,S^{J\star}_{bp}
\right)\,.
\nonumber
\end{eqnarray}
We can again replace $\sum_{a\neq b}$ by $\sum_{a,b}$, since the sum over $a=b$ cancels out in the difference, obtaining a product of Dirac deltas. Moreover, we recall that $m<n$ and $p<q$, so we can drop terms like $\delta_{mq}$ and $\delta_{np}$. Hence we have
\begin{equation}
(S^{(J,\psi)}\cdot S^{(J,\psi)\dagger})_{(mn)(pq)}=
\delta_{mp}\,\delta_{nq}-\delta_{mq}\,\delta_{np}=
\delta_{mp}\,\delta_{nq}\,.
\nonumber
\end{equation}
This is in agreement with unitarity.

\subsubsection*{Off-diagonal-twisted}
Compute:
\begin{eqnarray}
&& (S^{(J,\psi)}\cdot S^{(J,\psi)\dagger})_{(mn)\widehat{(p,\chi)}}= \nonumber\\
&&=
\sum_{(i,\xi)} S^{(J,\psi)}_{(mn)(i,\xi)}\,S^{(J,\psi)\dagger}_{(i,\xi)\widehat{(p,\chi)}}+ 
\sum_{(ab)} S^{(J,\psi)}_{(mn)(ab)}\,S^{(J,\psi)\dagger}_{(ab)\widehat{(p,\chi)}}+
\sum_{\widehat{(q,\xi)}} S^{(J,\psi)}_{(mn)\widehat{(q,\xi)}}\,S^{(J,\psi)\dagger}_{\widehat{(q,\xi)}\widehat{(p,\chi)}}=\nonumber\\
&&=
\sum_{\xi=0}^1 \sum_{i,Ji=i} S^J_{mi}\,S^J_{ni}\,
C^\star\,\frac{1}{2}\,S^{\star}_{ip}\,e^{-i\pi\xi}+\nonumber\\
&&+
\sum_{(a,b)} (S^J_{ma}\,S^J_{nb}+(-1)^\psi S^J_{mb}\,S^J_{na}) \cdot
\left\{
\begin{array}{cl}
0 & {\,\,\rm if \,\,} Ja=a \\
A^\star\,S_{ap}^\star & {\,\,\rm if \,\,} Ja=b
\end{array}
\right. +\nonumber\\
&&+
\sum_{\xi=0}^1 \sum_{q,\,\widehat{(q,\xi)}\,\,{\rm f.p.of\,\,}(J,\psi)} 
\left\{
\begin{array}{cl}
0 & {\,\,\rm if \,\,} Jm=m\\
A\,S_{mq} & {\,\,\rm if \,\,} Jm=n
\end{array}
\right.
\cdot
B^\star\,\frac{1}{2}\,e^{-i\pi\hat{Q}_J(q)}\,P^\star_{Jq,p}\,e^{i\pi(\xi+\chi)}=\nonumber\\
&&=
0\,+\,0\,+\,0\,=\,0\,,\nonumber
\end{eqnarray}
as it should be by unitarity. The first and third lines vanish because $\sum_{\xi=0}^1 e^{i\pi\xi}=0$. The second line vanishes when $(mn)$ has orbit-like form, because of $S^{J}=0$, but also if $m$ and $n$ are fixed points of $J$ by the ansatz.

\subsubsection*{Twisted-twisted}
Compute:
\begin{eqnarray}
&& (S^{(J,\psi)}\cdot S^{(J,\psi)\dagger})_{\widehat{(p,\phi)}\widehat{(q,\chi)}}= \nonumber\\
&&=
\sum_{(i,\xi)} S^{(J,\psi)}_{\widehat{(p,\phi)}(i,\xi)}\,S^{(J,\psi)\dagger}_{(i,\xi)\widehat{(q,\chi)}}+ 
\sum_{(mn)} S^{(J,\psi)}_{\widehat{(p,\phi)}(mn)}\,S^{(J,\psi)\dagger}_{(mn)\widehat{(q,\chi)}}+
\sum_{\widehat{(a,\xi)}} S^{(J,\psi)}_{\widehat{(p,\phi)}\widehat{(a,\xi)}}\,S^{(J,\psi)\dagger}_{\widehat{(a,\xi)}\widehat{(q,\chi)}}=
\nonumber\\
&&=
\sum_{\xi=0}^1 \sum_{i,Ji=i}
\frac{1}{2}\,S_{pi}\,e^{i\pi\xi}\,\frac{1}{2}S^\star_{iq}\,e^{-i\pi\xi}+\nonumber\\
&&+
\sum_{(mn)} 
\left\{
\begin{array}{cl}
0 & {\,\,\rm if \,\,} Jm=m \\
A\,S_{pm} & {\,\,\rm if \,\,} Jm=n
\end{array}
\right.
\cdot
\left\{
\begin{array}{cl}
0 & {\,\,\rm if \,\,} Jm=m\\
A^\star\,S_{mq}^\star & {\,\,\rm if \,\,} Jm=n
\end{array}
\right. +\nonumber\\
&&+
\sum_{\xi=0}^1 \sum_{a,\,\widehat{(a,\xi)}\,\,{\rm f.p.of\,\,}(J,\psi)} 
\frac{1}{2}\,e^{i\pi\hat{Q}_J(a)}\, P_{Ja,p}\,e^{i\pi(\phi+\xi)}\, \frac{1}{2}\,e^{-i\pi\hat{Q}_J(a)}\,P^\star_{Ja,q}\,e^{i\pi(\xi+\chi)}=\nonumber\\
&&=
\frac{1}{2}\,\sum_{i,Ji=i} S_{pi}\,S^\star_{iq}+\nonumber\\
&&+
\sum_{(mn), n=Jm} S_{pm}\,S^\star_{mq}+\nonumber\\
&&+
\frac{1}{2}\,e^{i\pi(\phi+\chi)}\,\sum_{a,\,\widehat{(a,\xi)}\,\,{\rm f.p.of\,\,}(J,\psi)} P_{Ja,p}\,P^\star_{Ja,q}\,.\nonumber
\end{eqnarray}
Here we have retained the second line of the off-diagonal-twisted ansatz, since only the off-diagonal fields with orbit-like form $(m,n\equiv Jm)$ contribute. We can rewrite the sum in the second line as
\begin{equation}
\sum_{(mn), n=Jm} S_{pm}\,S^\star_{mq}=\frac{1}{2}
\sum_{m} (S_{pm}\,S^\star_{mq} + 
S_{p,Jm}\,S^\star_{Jm,q})=
\frac{1}{2}\sum_{\forall m,Jm\neq m} S_{pm}\,S^\star_{mq}\,,\nonumber
\nonumber
\end{equation}
since $S_{p,Jm}=e^{2 i \pi Q_J(p)}S_{pm}$ and $S^\star_{Jm,q}=e^{-2 i \pi Q_J(q)}S_{mq}$, with $p$ and $q$ having the same monodromy charge and hence dropping out in the product. This combines then with the first line, while we can use our Corollary 2 in the third line. Hence:
\begin{eqnarray}
&& (S^{(J,\psi)}\cdot S^{(J,\psi)\dagger})_{\widehat{(p,\phi)}\widehat{(q,\chi)}}= \nonumber\\
&&=
\frac{1}{2}\,\underbrace{\left[\sum_{i,Ji=i}S_{pi}\,S^\star_{iq}+\sum_{m,Jm\neq m}S_{pm}\,S^\star_{mq}\right]}_{=(S\cdot S^\star)_{pq}=\delta_{pq}}
+\frac{1}{2} \,e^{i\pi(\phi+\chi)} \sum_{a,\,\widehat{(a,\xi)}\,\,{\rm f.p.of\,\,}(J,\psi)} P_{Ja,p}\,P^\star_{Ja,q}=\nonumber\\
&&=
\frac{1}{2}\,\delta_{pq}+\frac{1}{2}\,e^{i\pi(\phi+\chi)}\,\delta_{pq}=\delta_{pq}\,\delta_{\phi\chi}\,,
\nonumber
\end{eqnarray}
as needed for unitarity.

\newpage
\section{Modular invariance}
In this section we prove unitarity of the ansatz via the relation 
\begin{equation}
{T^{(J,\psi)}}^{-1} S^{(J,\psi)} {T^{(J,\psi)}}^{-1}=S^{(J,\psi)} T^{(J,\psi)} S^{(J,\psi)}\,,
\end{equation}
that must be checked sector by sector. Before doing this let us collect a few facts about $\mathbb{Z}_2$ permutation orbifolds that we will need in the calculation. The central charge $\hat{c}$ and the weights $\{h_{(mn)},\,h_{(i,\phi)},\,h_{\widehat{(i,\xi)}}\}$ are related to the analogous quantities of the mother CFT as
\begin{eqnarray}
\hat{c}&=&2\,c\\
h_{(mn)}&=&h_m+h_n\\
h_{(i,\phi)}&=&2\,h_i+\delta_{i,0}\,\delta_{\phi,1}\\
h_{\widehat{(i,\xi)}}&=&\frac{h_i}{2}+\frac{c}{16}+\frac{\xi}{2}\,.
\end{eqnarray}

Moreover, in the mother theory the $S^J$ matrix is unitary and modular invariant. Explicitly, the constraint $(T^{-1}S^JT^{-1})_{im}=(S^JTS^J)_{im}$ is
\begin{equation}
\label{constraint for mod inv}
\sum_{n,\,Jn=n} S^J_{in}\,e^{2 i\pi(h_n-\frac{c}{24})}\,S^J_{nm}=
e^{-2 i\pi(h_i-\frac{c}{24})}\,S^J_{im}\,e^{-2 i\pi(h_j-\frac{c}{24})}\,
\end{equation}
where of course $i$ and $j$ are fixed points of $J$.

Another quantity that it will be useful to spell out is $P_{Jp,q}$:
\begin{equation}
\label{P_(Jp,q)}
P_{Jp,q}=e^{i\pi(h_J-\hat{Q}_J(p))}\,\sqrt{T}_p\,
\sum_m e^{2 i\pi Q_J(m)}\,S_{pm}\,T^2_m\,S_{mq}\,\sqrt{T}_q\,,
\end{equation}
where $\hat{Q}_J(m)=h_J+h_m-h_{Jm}$. This comes from re-expressing $T_{Jp}$ in terms of $T_p$ as
\begin{equation}
T_{Jp}=T_p\cdot e^{2 i\pi (h_J-\hat{Q}_J(p))}\,.
\end{equation}

Now we are ready to check modular invariance. We will use the ansatz as given in (\ref{ansatz with fixed points}), keeping only $A$, $B$ and $C$ as undetermined phases and having fixed the others to one.

\subsubsection*{Diagonal-diagonal}
This matrix element exists only when $\psi=0$ and $i$ and $j$ are fixed points of $J$.
Compare the two expressions. On one side:
\begin{eqnarray}
&& \left({T^{(J,\psi)}}^{-1} S^{(J,\psi)} {T^{(J,\psi)}}^{-1}\right)_{(i,\phi)(j,\chi)}=
{T^{(J,\psi)}}^{-1}_{(i,\phi)} S^{(J,\psi)}_{(i,\phi)(j,\chi)} {T^{(J,\psi)}}^{-1}_{(j,\chi)}= \nonumber\\
&&=
e^{-2i\pi(h_{(i,\phi)}-\frac{\hat{c}}{24})}\,\frac{1}{2}\,S^J_{ij}\,S^J_{ij}
\,e^{-2i\pi(h_{(j,\chi)}-\frac{\hat{c}}{24})}=\nonumber\\
&&=
\frac{1}{2}\, T^{-2}_i \,S^J_{ij}\,S^J_{ij}\,T^{-2}_j \,.\nonumber
\end{eqnarray}
On the other side:
\begin{eqnarray}
&& \left(S^{(J,\psi)} T^{(J,\psi)} S^{(J,\psi)}\right)_{(i,\phi)(j,\chi)}=
\sum_{(a,\xi)} S^{(J,\psi)}_{(i,\phi)(a,\xi)} T^{(J,\psi)}_{(a,\xi)} S^{(J,\psi)}_{(a,\xi)(j,\chi)}+
\nonumber\\
&&+ 
\sum_{(mn)} S^{(J,\psi)}_{(i,\phi)(mn)} T^{(J,\psi)}_{(mn)} S^{(J,\psi)}_{(mn)(j,\chi)}+ 
\sum_{\widehat{(a,\xi)}} S^{(J,\psi)}_{(i,\phi)\widehat{(a,\xi)}} T^{(J,\psi)}_{\widehat{(a,\xi)}} S^{(J,\psi)}_{\widehat{(a,\xi)}(j,\chi)}=\nonumber\\
&&=
\sum_{\xi=0}^1 \sum_{a,\,Ja=a} \frac{1}{2}\,S^J_{ia}\,S^J_{ia}\, e^{2i\pi(2\,h_a-\frac{c}{12})}\,
\frac{1}{2}\,S^J_{aj}\,S^J_{aj}+\nonumber\\
&&+
\sum_{(mn)} S^J_{im}\,S^J_{in}\, e^{2i\pi(h_m +h_n-\frac{c}{12})}\, S^J_{mj}\,S^J_{nj}+\nonumber\\
&&+
\sum_{\xi=0}^1 \sum_{a,\,\widehat{(a,\xi)}\,\,{\rm f.p.of\,\,}(J,\psi)} C^2\,
\frac{1}{2}\,S_{ia}\,e^{i\pi\phi}\,e^{2i\pi(\frac{h_a}{2}+\frac{\xi}{2}-\frac{c}{48})}\,
\frac{1}{2}\,S_{aj}\,e^{i\pi\chi}\nonumber\,.
\end{eqnarray}
Now the last line vanishes, due to $\sum_{\xi=0}^1 e^{i\pi\xi}=0$. In the second line, $\sum_{(mn)}$ is restricted to $m<n$ with $m$ and $n$ fixed points of $J$ and can be replaced by $\sum_{(mn)}=\frac{1}{2}\sum_{m,n\,m\neq n}$, while the first line completes the sum over $m=n$, $\sum_a\rightarrow \sum_{m=n}$. Hence the first two lines combine to give:
\begin{eqnarray}
&& \left(S^{(J,\psi)} T^{(J,\psi)} S^{(J,\psi)}\right)_{(i,\phi)(j,\chi)}=\dots=\nonumber\\
&&=
\frac{1}{2}\,\sum_m\sum_n S^J_{im}\,S^J_{in}\,S^J_{mj}\,S^J_{nj}\,
e^{2i\pi(h_m+h_n-\frac{c}{12})}=\nonumber\\
&&=
\frac{1}{2}\,\left[T^{-1}_i\,S^J_{ij}\,T^{-1}_j\right]^2 \nonumber\,
\end{eqnarray}
after using (\ref{constraint for mod inv}). Hence:
\begin{equation}
\left(S^{(J,\psi)} T^{(J,\psi)} S^{(J,\psi)}\right)_{(i,\phi)(j,\chi)}=
\left({T^{(J,\psi)}}^{-1} S^{(J,\psi)} {T^{(J,\psi)}}^{-1}\right)_{(i,\phi)(j,\chi)}\,.
\end{equation}

\subsubsection*{Diagonal-off-diagonal}
This matrix elements only exists for $\psi=0$. Moreover, it would be clear from the calculation that we need to consider only the case when $(mn)$ is fixed-point-like, i.e. when $m$ and $n$ are fixed points of $J$. In the remaining case when $(mn)$ is orbit-like, modular invariance is trivially satisfied as $0\equiv 0$, since in that case $S^J$ vanishes.

Consider first:
\begin{eqnarray}
&& = \left({T^{(J,\psi)}}^{-1} S^{(J,\psi)} {T^{(J,\psi)}}^{-1}\right)_{(i,\phi)(mn)}=
{T^{(J,\psi)}}^{-1}_{(i,\phi)} S^{(J,\psi)}_{(i,\phi)(mn)} {T^{(J,\psi)}}^{-1}_{(mn)}= \nonumber\\
&&=
e^{-2i\pi(2\,h_i-\frac{c}{12})}\,S^J_{im}\,S^J_{in}\, e^{-2i\pi(h_m+h_n-\frac{c}{12})}=\nonumber\\
&&=
T^{-2}_i \,S^J_{im}\,S^J_{in}\,T^{-1}_m\,T^{-1}_n \,.\nonumber
\end{eqnarray}
Then, consider
\begin{eqnarray}
&& \left(S^{(J,\psi)} T^{(J,\psi)} S^{(J,\psi)}\right)_{(i,\phi)(mn)}=
\sum_{(a,\xi)} S^{(J,\psi)}_{(i,\phi)(a,\xi)} T^{(J,\psi)}_{(a,\xi)} S^{(J,\psi)}_{(a,\xi)(mn)}+
\nonumber\\
&&+ 
\sum_{(ab)} S^{(J,\psi)}_{(i,\phi)(ab)} T^{(J,\psi)}_{(ab)} S^{(J,\psi)}_{(ab)(mn)}+ 
\sum_{\widehat{(a,\xi)}} S^{(J,\psi)}_{(i,\phi)\widehat{(a,\xi)}} T^{(J,\psi)}_{\widehat{(a,\xi)}} S^{(J,\psi)}_{\widehat{(a,\xi)}(mn)}=\nonumber\\
&&=
\sum_{\xi=0}^1 \sum_{a,\,Ja=a} 
\frac{1}{2}\,S^J_{ia}\,S^J_{ia}\, e^{2i\pi(2\,h_a-\frac{c}{12})}\, S^J_{am}\,S^J_{an}+\nonumber\\
&&+
\sum_{(ab)} S^J_{ia}\,S^J_{ib}\, e^{2i\pi(h_a +h_b-\frac{c}{12})}\, 
\left(S^J_{am}\,S^J_{bn}+ (-1)^\psi S^J_{an}\,S^J_{bm}\right)+\nonumber\\
&&+
\sum_{\xi=0}^1 \sum_{a,\,Ja=a} C\,
\frac{1}{2}\,S_{ia}\,e^{i\pi\phi}\,e^{2i\pi(\frac{h_a}{2}+\frac{\xi}{2}-\frac{c}{48})}\,\cdot
\left\{
\begin{array}{cl}
0 & {\,\,\rm if\,\,} Jm=m\\
A\,S_{mp} & {\,\,\rm if\,\,} Jm=n
\end{array}
\right.\nonumber\,.
\end{eqnarray}
Now, the last line vanishes by the ansatz if $Jm=m$ and because of $\sum_\xi e^{i\pi\xi}=0$ if $Jm=n$. The sum over $(ab)$ is restricted to the fixed points of $J$ and to $a<b$. recalling that $\psi=0$, the two terms in that sum allows us to rewrite it as a sum over all $a$ and $b$ with $a\neq b$, i.e.:
\begin{equation}
\sum_{(ab)} S^J_{ia}\,S^J_{ib}\, T_a\,T_b\,\left(S^J_{am}\,S^J_{bn}+ S^J_{an}\,S^J_{bm}\right)=
\sum_{a,b,\,a\neq b} S^J_{ia}\,S^J_{ib}\, T_a\,T_b\,\, S^J_{am}\,S^J_{bn}\,.
\nonumber
\end{equation}
Hence we can combine the first two lines and then use the constraint (\ref{constraint for mod inv}) to get:
\begin{eqnarray}
&& \left(S^{(J,\psi)} T^{(J,\psi)} S^{(J,\psi)}\right)_{(i,\phi)(mn)}=\dots=\nonumber\\
&&=
\sum_a\sum_b  e^{2i\pi(h_a +h_b-\frac{c}{12})}\, S^J_{ia}\,S^J_{ib}\,S^J_{am}\,S^J_{bn}=\nonumber\\
&&=T^{-1}_i\,S^J_{im}\,T^{-1}_m\,T^{-1}_i\,S^J_{in}\,T^{-1}_n\,.\nonumber
\end{eqnarray}
Hence:
\begin{equation}
\left(S^{(J,\psi)} T^{(J,\psi)} S^{(J,\psi)}\right)_{(i,\phi)(mn)}=
\left({T^{(J,\psi)}}^{-1} S^{(J,\psi)} {T^{(J,\psi)}}^{-1}\right)_{(i,\phi)(mn)}\,.
\end{equation}

\subsubsection*{Off-diagonal-off-diagonal}
This matrix elements is always present in any extension of permutation orbifolds. In order to check modular invariance we need to consider only fixed-point-like off-diagonal fields. In fact, since $S^J$ matrices of the original theory are involved, the constraint (\ref{constraint for mod inv}) is trivially satisfied, in the form $0\equiv 0$, when at least one off-diagonal field is orbit-like. However, in the sum over all fields of the extended permutation orbifolds, the diagonal fields appear only for $\psi=0$.

Compute:
\begin{eqnarray}
&& \left({T^{(J,\psi)}}^{-1} S^{(J,\psi)} {T^{(J,\psi)}}^{-1}\right)_{(mn)(pq)}=
{T^{(J,\psi)}}^{-1}_{(mn)} S^{(J,\psi)}_{(mn)(pq)} {T^{(J,\psi)}}^{-1}_{(pq)}= \nonumber\\
&&=
e^{-2i\pi(h_m+h_n-\frac{c}{12})}\,\left(S^J_{mp}\,S^J_{nq}+(-1)^\psi\,S^J_{mq}\,S^J_{np}\right)\,
e^{-2i\pi(h_p+h_q-\frac{c}{12})}=\nonumber\\
&&=
T^{-1}_m\,T^{-1}_n\,\left(S^J_{mp}\,S^J_{nq}+(-1)^\psi\,S^J_{mq}\,S^J_{np}\right)\,T^{-1}_p\,T^{-1}_q \,.\nonumber
\end{eqnarray}
The analogous quantity is:
\begin{eqnarray}
&& \left(S^{(J,\psi)} T^{(J,\psi)} S^{(J,\psi)}\right)_{(mn)(pq)}=
\delta_{\psi,0}\,\sum_{(a,\xi)} S^{(J,\psi)}_{(mn)(a,\xi)} T^{(J,\psi)}_{(a,\xi)} S^{(J,\psi)}_{(a,\xi)(pq)}+
\nonumber\\
&&+ 
\sum_{(ab)} S^{(J,\psi)}_{(mn)(ab)} T^{(J,\psi)}_{(ab)} S^{(J,\psi)}_{(ab)(pq)}+ 
\sum_{\widehat{(a,\xi)}} S^{(J,\psi)}_{(mn)\widehat{(a,\xi)}} T^{(J,\psi)}_{\widehat{(a,\xi)}} S^{(J,\psi)}_{\widehat{(a,\xi)}(pq)}=\nonumber\\
&&=
\delta_{\psi,0}\,\sum_{\xi=0}^1 \sum_{a,\,Ja=a} 
S^J_{ma}\,S^J_{na}\,e^{2i\pi(2\,h_a-\frac{c}{12})}\, S^J_{ap}\,S^J_{aq}+\nonumber\\
&&+
\sum_{(ab)} \left(S^J_{ma}\,S^J_{nb}+(-1)^\psi\,S^J_{mb}\,S^J_{na}\right)\, e^{2i\pi(h_a +h_b-\frac{c}{12})}\, 
\left(S^J_{ap}\,S^J_{bq}+ (-1)^\psi S^J_{aq}\,S^J_{bp}\right)+\nonumber\\
&&+
\sum_{\xi=0}^1 \sum_{a,\,\widehat{(a,\xi)}\,\,{\rm f.p.of\,\,}(J,\psi)} \nonumber\\
&&
\phantom{\sum_{\xi=0}^1}
\left\{
\begin{array}{cl}
0 & {\,\,\rm if\,\,} Jm=m\\
A\,S_{ma}  & {\,\,\rm if\,\,} Jm=n
\end{array}
\right.
e^{2i\pi(\frac{h_a}{2}+\frac{\xi}{2}-\frac{c}{48})}
\left\{
\begin{array}{cl}
0 & {\,\,\rm if\,\,} Jp=q\\
A\,S_{ap}  & {\,\,\rm if\,\,} Jp=q
\end{array}
\right.\,.\nonumber
\end{eqnarray}
The last line always vanishes, either when $Jm=m$ by the ansatz or when $Jm=n$ because of $\sum_{\xi=0}^1 e^{i\pi\xi}=0$. The second line can be splitted in two pieces and rewritten as
\begin{eqnarray}
&&\sum_{(ab)} \left(S^J_{ma}\,S^J_{nb}+(-1)^\psi\,S^J_{mb}\,S^J_{na}\right)\, e^{2i\pi(h_a +h_b-\frac{c}{12})}\, 
\left(S^J_{ap}\,S^J_{bq}+ (-1)^\psi S^J_{aq}\,S^J_{bp}\right)=\nonumber\\
&&=
\sum_{a\neq b} S^J_{ma}\,S^J_{nb}\,S^J_{ap}\,S^J_{bq}\,e^{2i\pi(h_a +h_b-\frac{c}{12})} +
(-1)^\psi\sum_{a\neq b} S^J_{mb}\,S^J_{na}\,S^J_{ap}\,S^J_{bq}\,e^{2i\pi(h_a +h_b-\frac{c}{12})}\nonumber
\end{eqnarray}
where now we can replace $\sum_{a\neq b}\rightarrow \sum_a \sum_b -\sum_{a=b}$. This leave us with two sums over $a$ and $b$ plus three equal sums over $a$, one with a factor $2\,\delta_{\psi,0}$ in front, one with a factor $-1$ and one with $-(-1)^\psi$, whose  combination vanishes:
\begin{equation}
\underbrace{(2\,\delta_{\psi,0}-1-(-1)^\psi)}_{=0}\,\cdot\sum_a S^J_{ma}\,S^J_{na}\,S^J_{ap}\,S^J_{aq}\,e^{2i\pi(2\,h_a-\frac{c}{12})}=0 \nonumber
\end{equation}
We can use (\ref{constraint for mod inv}) in the two remaining contributions and get:
\begin{eqnarray}
&& \left(S^{(J,\psi)} T^{(J,\psi)} S^{(J,\psi)}\right)_{(mn)(pq)}=\dots=\nonumber\\
&&=
\sum_a S^J_{ma}\,T_a\,S^J_{ap}\cdot\sum_b S^J_{nb}\,T_b\,S^J_{bq}+
(-1)^\psi \sum_a S^J_{na}\,T_a\,S^J_{ap}\cdot\sum_b S^J_{mb}\,T_b\,S^J_{bq}=\nonumber\\
&&=
T^{-1}_m\,S^J_{mp}\,T^{-1}_{p}\cdot T^{-1}_n\,S^J_{nq}\,T^{-1}_{q}+
(-1)^\psi T^{-1}_n\,S^J_{np}\,T^{-1}_{p}\cdot T^{-1}_m\,S^J_{mq}\,T^{-1}_{q}=\nonumber\\
&&=
T^{-1}_m\,T^{-1}_n\,\left(S^J_{mp}\,S^J_{nq}+(-1)^\psi S^J_{np}\,S^J_{mq}\right)\, T^{-1}_{p}\,T^{-1}_{q}\,.\nonumber
\end{eqnarray}
Hence:
\begin{equation}
\left(S^{(J,\psi)} T^{(J,\psi)} S^{(J,\psi)}\right)_{(mn)(pq)}=
\left({T^{(J,\psi)}}^{-1} S^{(J,\psi)} {T^{(J,\psi)}}^{-1}\right)_{(mn)(pq)}\,.
\end{equation}

\subsubsection*{Twisted-twisted}
Compute:
\begin{eqnarray}
&& \left({T^{(J,\psi)}}^{-1} S^{(J,\psi)} {T^{(J,\psi)}}^{-1}\right)_{\widehat{(p,\phi)}\widehat{(q,\chi)}}=
{T^{(J,\psi)}}^{-1}_{\widehat{(p,\phi)}} S^{(J,\psi)}_{\widehat{(p,\phi)}\widehat{(q,\chi)}} {T^{(J,\psi)}}^{-1}_{\widehat{(q,\chi)}}= \nonumber\\
&&=
e^{-2i\pi(\frac{h_p}{2}+\frac{\phi}{2}-\frac{c}{48})}\,B\,\frac{1}{2}\,e^{i\pi\hat{Q}_J(p)}\,P_{Jp,q}\, e^{i\pi(\phi+\chi)}\, e^{-2i\pi(\frac{h_q}{2}+\frac{\chi}{2}-\frac{c}{48})}=\nonumber\\
&&=
\frac{1}{2}\, B\,e^{i\pi\hat{Q}_J(p)}\,\sqrt{T}^{-1}_p \,P_{Jp,q}\,\sqrt{T}^{-1}_q \,.\nonumber
\end{eqnarray}
On the other hand we have
\begin{eqnarray}
&& \left(S^{(J,\psi)} T^{(J,\psi)} S^{(J,\psi)}\right)_{\widehat{(p,\phi)}\widehat{(q,\chi)}}=
\delta_{\psi,0}\,\sum_{(a,\xi)} S^{(J,\psi)}_{\widehat{(p,\phi)}(a,\xi)} T^{(J,\psi)}_{(a,\xi)} S^{(J,\psi)}_{(a,\xi)\widehat{(q,\chi)}}+
\nonumber\\
&&+ 
\sum_{(ab)} S^{(J,\psi)}_{\widehat{(p,\phi)}(ab)} T^{(J,\psi)}_{(ab)} S^{(J,\psi)}_{(ab)\widehat{(q,\chi)}}+ 
\sum_{\widehat{(a,\xi)}} S^{(J,\psi)}_{\widehat{(p,\phi)}\widehat{(a,\xi)}} T^{(J,\psi)}_{\widehat{(a,\xi)}} S^{(J,\psi)}_{\widehat{(a,\xi)}\widehat{(q,\chi)}}=\nonumber\\
&&=
\delta_{\psi,0}\,\sum_{\xi=0}^1 \sum_{a,\,Ja=a} C^2 \frac{1}{2}\,S_{pa}\,e^{i\pi\xi}\, 
e^{2i\pi(2\,h_a-\frac{c}{12})}\,\frac{1}{2}\,S_{aq}\,e^{i\pi\xi}+\nonumber\\
&&+
\sum_{(ab)}
\left\{
\begin{array}{cl}
0 & {\,\,\rm\,\,} Ja=a\\
A\,S_{pa} & {\,\,\rm\,\,} Ja=b
\end{array}
\right. 
\cdot
e^{2i\pi(h_a+h_b-\frac{c}{12})}
\cdot
\left\{
\begin{array}{cl}
0 & {\,\,\rm\,\,} Ja=a\\
A\,S_{aq} & {\,\,\rm\,\,} Ja=b
\end{array}
\right. +\nonumber\\
&&+
\sum_{\xi=0}^1 \sum_{a,\,\widehat{(a,\xi)}\,\,{\rm f.p.of\,\,}(J,\psi)} B^2\,
\frac{1}{2}\,e^{i\pi\hat{Q}_J(a)}\, P_{p,Ja}\,e^{i\pi(\phi+\xi)} \,e^{2i\pi(\frac{h_a}{2}+\frac{\xi}{2}-\frac{c}{48})}\,
\frac{1}{2}\,e^{i\pi\hat{Q}_J(a)}\,P_{Ja,q}\,e^{i\pi(\xi+\chi)}\nonumber\,.
\end{eqnarray}
Now the last line vanishes because $\sum_{\xi=0}^1 e^{i\pi\xi}=0$. So we have
\begin{eqnarray}
\label{tw-tw temporary expression}
&& \left(S^{(J,\psi)} T^{(J,\psi)} S^{(J,\psi)}\right)_{\widehat{(p,\phi)}\widehat{(q,\chi)}}=\dots=\\
&&=
\frac{1}{2}\, \left[
\delta_{\psi,0}\,\sum_{a,\,Ja=a} C^2\,
S_{pa}\,T^2_a\,S_{aq}+
\sum_{a,\,Ja\neq a} A^2\,
S_{pa}\,T_a\,T_{Ja}\,S_{aq}
\right]\nonumber
\end{eqnarray}

The two contributions combine nicely, provided we take $A^2=C^2$. We will actually prove in a moment that this must be indeed the case (at least when $\psi=0$). Since for a fixed point $a$, $\hat{Q}_J(a)=h_J$, we can always either write or cancel a factor $e^{2i\pi (h_J-\hat{Q}_J(a))}$ in the sum over the fixed points. Moreover, in the second term we can replace the sum over non-fixed-points with the sum over all the fields minus the sum over the fixed points:
\begin{equation}
A^2\,\sum_{a,\,Ja\neq a}=A^2\,\sum_a -A^2\,\sum_{a,\,Ja=a}\,. \nonumber
\end{equation}
From the case $\psi=0$, the prefactor $A^2$ will be set equal to $C^2$, while for $\psi=1$ all the unwanted contributions will cancel out.

So we have
\begin{eqnarray}
&& \left(S^{(J,\psi)} T^{(J,\psi)} S^{(J,\psi)}\right)_{\widehat{(p,\phi)}\widehat{(q,\chi)}}=\dots=\nonumber\\
&&=
\frac{1}{2}\, \left[
\delta_{\psi,0}\,\sum_{a,\,Ja=a} C^2\,
S_{pa}\,T^2_a\,S_{aq}+\right.\nonumber\\
&&
\phantom{\frac{1}{2}\left[\right.\,}\,
+
\left. \sum_{\forall a} A^2\,
S_{pa}\,T^2_a\,e^{2i\pi(h_J-\hat{Q}_J(a))}\,S_{aq}-
\sum_{a,\,Ja=a} A^2\,S_{pa}\,T^2_a\,S_{aq}
\right]=\nonumber\\
&&=
\frac{1}{2}\, \left[
(\delta_{\psi,0}\,C^2-A^2)\,\sum_{a,\,Ja=a}
S_{pa}\,T^2_a\,S_{aq}
+\sum_{a} A^2\,S_{pa}\,T^2_a\,e^{2i\pi(h_J-\hat{Q}_J(a))}\,S_{aq}\right]\,.\nonumber
\end{eqnarray}
In the second contribution we can use (\ref{P_(Jp,q)}), replacing $-\hat{Q}_J(a)$ by $Q_J(a)$ in the exponent, since they differ only by integers and $Q_J(a)$ is either $0$ or $\frac{1}{2}$. This will give agreement with $\left({T^{(J,\psi)}}^{-1} S^{(J,\psi)} {T^{(J,\psi)}}^{-1}\right)_{\widehat{(p,\phi)}\widehat{(q,\chi)}}$. Hence the first contribution must vanish and indeed it does so, by requiring that $C^2-A^2=0$ for $\psi=0$, while for $\psi=1$ it vanishes automatically, since $S_{pa}=S_{aq}=0$ for (half-)integer spin simple currents\footnote{Strictly speaking, this is not true when $h_J\in\frac{1}{4}\, \mathbb{Z}_{\rm odd}$, since we would then naively have $S_{pa}=(-1)^{\psi+4h_J}\,S_{pa}$, but in that case $J$ has no fixed points, hence there is no such a sum in the calculation.}, due to (\ref{S=0 if Jm=m and psi=1}). In the end we are left with
\begin{eqnarray}
&& \left(S^{(J,\psi)} T^{(J,\psi)} S^{(J,\psi)}\right)_{\widehat{(p,\phi)}\widehat{(q,\chi)}}=\dots=\nonumber\\
&&=
\frac{1}{2}\,A^2\,e^{2i\pi h_J}
\sum_{a} \,e^{2i\pi Q_J(a)}\,S_{pa}\,T^2_a\,S_{aq}=\nonumber\\
&&=
\frac{1}{2}\,A^2\,e^{i\pi h_J}\,e^{i\pi \hat{Q}_J(p)}\,
\sqrt{T}^{-1}_p\,P_{Jp,q}\,\sqrt{T}^{-1}_q\,.\nonumber
\end{eqnarray}
Hence:
\begin{equation}
\left(S^{(J,\psi)} T^{(J,\psi)} S^{(J,\psi)}\right)_{\widehat{(p,\phi)}\widehat{(q,\chi)}}=
\left({T^{(J,\psi)}}^{-1} S^{(J,\psi)} {T^{(J,\psi)}}^{-1}\right)_{\widehat{(p,\phi)}\widehat{(q,\chi)}}\,,
\end{equation}
provided
\begin{equation}
\boxed{
B=A^2\,e^{i\pi h_J}\qquad {\rm and} \qquad C^2=A^2}\,.
\end{equation}
Here an observation is in order. Strictly speaking, the equality $C^2=A^2$ holds true only when $\psi=0$. When instead $\psi=1$, $C^2$ can in principle be different from $A^2$, since in this case we do not find any constraint on it. On the other side, however, when $\psi=1$ the value of $C$ is irrelevant, since $C$ enters the definition of the matrix element $S^{(J,\psi)}_{(i,\phi)\widehat{(p,\chi)}}$ but there are no diagonal fixed points $(i,\phi)$ in this case. Also note that we could have obtained this same result if we had replaced
\begin{equation}
C^2\,\sum_{a,\,Ja= a}=C^2\,\sum_a -\,C^2\,\sum_{a,\,Ja\neq a}\,. \nonumber
\end{equation}
in the intermediate eq. (\ref{tw-tw temporary expression}).

\subsubsection*{Diagonal-twisted}
This matrix element exists again only if $\psi=0$, otherwise there would survive no diagonal fields in the extension. Moreover, $h_J$ can be only (half-)integer, since for $h_J\in\frac{1}{4}\,\mathbb{Z}_{\rm odd}$ there are no fixed points in the mother theory.

Compute:
\begin{eqnarray}
&& \left({T^{(J,\psi)}}^{-1} S^{(J,\psi)} {T^{(J,\psi)}}^{-1}\right)_{(i,\phi)\widehat{(p,\chi)}}=
{T^{(J,\psi)}}^{-1}_{(i,\phi)} S^{(J,\psi)}_{(i,\phi)\widehat{(p,\chi)}} {T^{(J,\psi)}}^{-1}_{\widehat{(p,\chi)}}= \nonumber\\
&&=
e^{-2i\pi(2\,h_i-\frac{c}{12})}\,C\,\frac{1}{2}\,S_{ip}\, e^{i\pi\phi}\,e^{-2i\pi(\frac{h_p}{2}+\frac{\chi}{2}-\frac{c}{48})}=\nonumber\\
&&=
\frac{1}{2}\, C\,T^{-2}_i \,S_{ip}\,\sqrt{T}^{-1}_p \,e^{i\pi(\phi+\chi)}\,.\nonumber
\end{eqnarray}
On the other hand we have
\begin{eqnarray}
&& \left(S^{(J,\psi)} T^{(J,\psi)} S^{(J,\psi)}\right)_{(i,\phi)\widehat{(p,\chi)}}=
\sum_{(a,\xi)} S^{(J,\psi)}_{(i,\phi)(a,\xi)} T^{(J,\psi)}_{(a,\xi)} S^{(J,\psi)}_{(a,\xi)\widehat{(p,\chi)}}+
\nonumber\\
&&+ 
\sum_{(ab)} S^{(J,\psi)}_{(i,\phi)(ab)} T^{(J,\psi)}_{(ab)} S^{(J,\psi)}_{(ab)\widehat{(p,\chi)}}+ 
\sum_{\widehat{(a,\xi)}} S^{(J,\psi)}_{(i,\phi)\widehat{(a,\xi)}} T^{(J,\psi)}_{\widehat{(a,\xi)}} S^{(J,\psi)}_{\widehat{(a,\xi)}\widehat{(p,\chi)}}=\nonumber\\
&&=
\sum_{\xi=0}^1 \sum_{a,\,Ja=a} \frac{1}{2}\,S^J_{ia}\,S^J_{ia}\, e^{2i\pi(2\,h_a-\frac{c}{12})}\,
C\,\frac{1}{2}\,S_{ap}\,e^{i\pi\xi}+\nonumber\\
&&+
\sum_{(ab)} S^J_{ia}\,S^J_{ib}\, e^{2i\pi(h_a +h_b-\frac{c}{12})}\,
\cdot
\left\{
\begin{array}{cl}
0 & {\,\,\rm\,\,} Ja=a\\
A\,S_{ap} & {\,\,\rm\,\,} Ja=b
\end{array}
\right. +\nonumber\\
&&+
\sum_{\xi=0}^1 \sum_{a,\,\widehat{(a,\xi)}\,\,{\rm f.p.of\,\,}(J,\psi)} C\,
\frac{1}{2}\,S_{ia}\,e^{i\pi\phi}\,e^{2i\pi(\frac{h_a}{2}+\frac{\xi}{2}-\frac{c}{48})}\,
B\,\frac{1}{2}\,e^{i\pi\hat{Q}_J(a)}\,P_{Ja,p}\,\,e^{i\pi(\xi+\chi)}\nonumber\,.
\end{eqnarray}
Now the first line vanishes because $\sum_{\xi=0}^1 e^{i\pi\xi}=0$. The second line vanishes as well, either because of the ansatz if $Ja=a$ or because $S^J=0$ if $Ja=b$. Only the third line survives. There we can use (\ref{P_(Jp,q)}) and write
\begin{eqnarray}
&& \left(S^{(J,\psi)} T^{(J,\psi)} S^{(J,\psi)}\right)_{(i,\phi)\widehat{(p,\chi)}}=\dots=\nonumber\\
&&=
\frac{1}{2}\,B\,C\,e^{i\pi(\phi+\chi)} \sum_{a,\,\widehat{(a,\xi)}\,\,{\rm f.p.of\,\,}(J,\psi)}
S_{ia}\,\sqrt{T}_a\,e^{i\pi h_J}\,\sqrt{T_a}\,\sum_m e^{2i\pi Q_J(m)}\,S_{am}\,T^2_m\,S_{mp}\,\sqrt{T}_p=\nonumber\\
&&=
\frac{1}{2}\,B\,C\,e^{i\pi(\phi+\chi)} \sum _m 
\left( \sum_{a,\,\widehat{(a,\xi)}\,\,{\rm f.p.of\,\,}(J,\psi)}
S_{ia}\,T_a\,S_{am}\right)
\,e^{i\pi h_J}\,e^{2i\pi Q_J(m)}\,T^2_m\,S_{mp}\,\sqrt{T}_p\,.\nonumber
\end{eqnarray}
Consider the quantity within squared brackets. We can remove the projection over twisted fields only and extend the sum over all $a$ by introducing the projector (\ref{projector PI}) as done in the calculation for unitarity:
\begin{eqnarray}
&& \left(\sum_{a,\,\widehat{(a,\xi)}\,\,{\rm f.p.of\,\,}(J,\psi)}
S_{ia}\,T_a\,S_{am}\right)=
\sum_a \Pi(a)\,S_{ia}\,T_a\,S_{am}=\nonumber\\
&&=
\sum_a \frac{1}{2}\,\sum_{n=0}^{1}\,e^{2 i\pi n (Q_J(p)-\frac{\psi}{2})}\,S_{ia}\,T_a\,S_{am}=
\frac{1}{2}\,\sum_{n=0}^{1}\,e^{-2 i\pi n \frac{\psi}{2}}\,\sum_a S_{J^n i,a}\,T_a\,S_{am}=\nonumber\\
&&=
\frac{1}{2}\,\sum_{n=0}^{1}\,e^{-2 i\pi n \frac{\psi}{2}}\,\sum_a S_{ia}\,T_a\,S_{am}=
\frac{1+(-1)^\psi}{2}\,\sum_a S_{ia}\,T_a\,S_{am}=\nonumber\\
&&=
\sum_a S_{ia}\,T_a\,S_{am}=(STS)_{im}=(T^{-1}ST^{-1})_{im}\,,\nonumber
\end{eqnarray}
where we have used the fact that $i$ is a fixed point of $J$, $Ji=i$, and that $\psi=0$. The last equality follows from modular invariance of the $S$ matrix of the original theory. Going back to our main expression we have then
\begin{eqnarray}
&& \left(S^{(J,\psi)} T^{(J,\psi)} S^{(J,\psi)}\right)_{(i,\phi)\widehat{(p,\chi)}}=\dots=\nonumber\\
&&=
\frac{1}{2}\,B\,C\,e^{i\pi(\phi+\chi)}\,e^{i\pi h_J}\, \sum_m (STS)_{im}\,e^{2i\pi Q_J(m)}\,T^2_m\,S_{mp}\,\sqrt{T}_p=\nonumber\\
&&=
\frac{1}{2}\,B\,C\,e^{i\pi(\phi+\chi)}\,e^{i\pi h_J}\, \sum_m (T^{-1}ST^{-1})_{im}\,e^{2i\pi Q_J(m)}\,T^2_m\,S_{mp}\,\sqrt{T}_p=\nonumber\\
&&=
\frac{1}{2}\,B\,C\,e^{i\pi(\phi+\chi)}\,e^{i\pi h_J}\, \sum_m T^{-1}_i\,S_{im}\,T^{-1}_m\,
e^{2i\pi Q_J(m)}\,T^2_m\,S_{mp}\,\sqrt{T}_p=\nonumber\\
&&=
\frac{1}{2}\,B\,C\,e^{i\pi(\phi+\chi)}\,e^{i\pi h_J}\, \sum_m T^{-1}_i\,S_{Ji,m}\,T^{-1}_m\,T^2_m\,S_{mp}\,\sqrt{T}_p=\nonumber\\
&&=
\frac{1}{2}\,B\,C\,e^{i\pi(\phi+\chi)}\,e^{i\pi h_J}\, T^{-1}_i\,\sum_m (S_{im}\,T^{-1}_m\,T^2_m\,S_{mp})\,\sqrt{T}_p=\nonumber\\
&&=
\frac{1}{2}\,B\,C\,e^{i\pi(\phi+\chi)}\,e^{i\pi h_J}\, T^{-1}_i\,(T^{-1}ST^{-1})_{ip}\,\sqrt{T}_p=\nonumber\\
&&=
\frac{1}{2}\,B\,C\,e^{i\pi(\phi+\chi)}\,e^{i\pi h_J}\, T^{-2}_i\,S_{ip}\,\sqrt{T}^{-1}_p\,.\nonumber
\end{eqnarray}
Hence:
\begin{equation}
\left(S^{(J,\psi)} T^{(J,\psi)} S^{(J,\psi)}\right)_{(i,\phi)\widehat{(p,\chi)}}=
\left({T^{(J,\psi)}}^{-1} S^{(J,\psi)} {T^{(J,\psi)}}^{-1}\right)_{(i,\phi)\widehat{(p,\chi)}}\,,
\end{equation}
provided
\begin{equation}
B=e^{-i\pi h_J}=e^{3i\pi h_J}\,.
\end{equation}
Recall that we have used here $\psi=0$ and $h_J$ (half-)integer. Actually we will show in the next subsection that the correct expression for $B$ is
\begin{equation}
B=(-1)^\psi\,e^{3i\pi h_J}\,,
\end{equation}
valid for any value of $\psi$ and also when $h_J\in\frac{1}{4}\,\mathbb{Z}_{\rm odd}$.

\subsubsection*{Off-diagonal-twisted}
In the following calculation we will consider the off-diagonal field $(mn)$ to be orbit-like, i.e. $Jm=n$. In the other case, i.e. when $m$ and $n$ are fixed points of $J$, the constraint reduces to the trivial identity $0\equiv 0$. 

In fact, if $(mn)$ is fixed-point-like, on one side we have
\begin{eqnarray}
&& \left({T^{(J,\psi)}}^{-1} S^{(J,\psi)} {T^{(J,\psi)}}^{-1}\right)_{(mn)\widehat{(p,\chi)}}=
{T^{(J,\psi)}}^{-1}_{(mn)} S^{(J,\psi)}_{(mn)\widehat{(p,\chi)}} {T^{(J,\psi)}}^{-1}_{\widehat{(p,\chi)}}=0\,,\nonumber
\end{eqnarray}
since $S^{(J,\psi)}_{(mn)\widehat{(p,\chi)}}=0$ by the ansatz. On the other side we have
\begin{eqnarray}
&& \left(S^{(J,\psi)} T^{(J,\psi)} S^{(J,\psi)}\right)_{(mn)\widehat{(p,\chi)}}=
\delta_{\psi,0}\,\sum_{(a,\xi)} S^{(J,\psi)}_{(mn)(a,\xi)} T^{(J,\psi)}_{(a,\xi)} S^{(J,\psi)}_{(a,\xi)\widehat{(p,\chi)}}+
\nonumber\\
&&+ 
\sum_{(ab)} S^{(J,\psi)}_{(mn)(ab)} T^{(J,\psi)}_{(ab)} S^{(J,\psi)}_{(ab)\widehat{(p,\chi)}}+ 
\sum_{\widehat{(a,\xi)}} S^{(J,\psi)}_{(mn)\widehat{(a,\xi)}} T^{(J,\psi)}_{\widehat{(a,\xi)}} S^{(J,\psi)}_{\widehat{(a,\xi)}\widehat{(p,\chi)}}=\nonumber\\
&&=
\delta_{\psi,0}\,\sum_{\xi=0}^1 \sum_{a,\,Ja=a} S^J_{ma}\,S^J_{na}\, e^{2i\pi(2\,h_a-\frac{c}{12})}\,
C\,\frac{1}{2}\,S_{ap}\,e^{i\pi\xi}+\nonumber\\
&&+
\sum_{(ab)} \left(S^J_{ma}\,S^J_{nb}+(-1)^\psi S^J_{mb}\,S^J_{na}\right)\, e^{2i\pi(h_a +h_b-\frac{c}{12})}\,
\cdot
\left\{
\begin{array}{cl}
0 & {\,\,\rm\,\,} Ja=a\\
A\,S_{ap} & {\,\,\rm\,\,} Ja=b
\end{array}
\right. +\nonumber\\
&&+
0\,=\,0\nonumber\,,
\end{eqnarray}
because each term vanishes individually, being $\sum_{\xi=0}^1 e^{i\pi\xi}=0$ in the first contribution, either $S^J_{ma}=0$ (if $(ab)$ is orbit-like) or $S^{(J,\psi)}_{(ab)\widehat{(p,\chi)}}=0$ (if $(ab)$ is fixed-point-like) in the second, and $S^{(J,\psi)}_{(mn)\widehat{(a,\xi)}}=0$ by the ansatz in the third.

Hence, from now on we can restrict ourselves to orbit-like off-diagonal fields $(mn)$. Compute:
\begin{eqnarray}
&& \left({T^{(J,\psi)}}^{-1} S^{(J,\psi)} {T^{(J,\psi)}}^{-1}\right)_{(mn)\widehat{(p,\chi)}}=
{T^{(J,\psi)}}^{-1}_{(mn)} S^{(J,\psi)}_{(mn)\widehat{(p,\chi)}} {T^{(J,\psi)}}^{-1}_{\widehat{(p,\chi)}}= \nonumber\\
&&=
e^{-2i\pi(h_m+h_n-\frac{c}{12})}\,A\,S_{mp}\,e^{i\pi\chi}\,
e^{-2i\pi(\frac{h_p}{2}+\frac{\chi}{2}-\frac{c}{48})}=\nonumber\\
&&=
A\,T^{-1}_m\,T^{-1}_n \,S_{mp}\,\sqrt{T}^{-1}_p\,e^{i\pi\chi} \,.\nonumber
\end{eqnarray}
Here we could trade $T_n$ with $T_m$ at the cost of introducing phases depending on $h_J$ and $\hat{Q}_J(m)$, but it will not be necessary. On the other side:
\begin{eqnarray}
&& \left(S^{(J,\psi)} T^{(J,\psi)} S^{(J,\psi)}\right)_{(mn)\widehat{(p,\chi)}}=
\delta_{\psi,0}\,\sum_{(a,\xi)} S^{(J,\psi)}_{(mn)(a,\xi)} T^{(J,\psi)}_{(a,\xi)} S^{(J,\psi)}_{(a,\xi)\widehat{(p,\chi)}}+
\nonumber\\
&&+ 
\sum_{(ab)} S^{(J,\psi)}_{(mn)(ab)} T^{(J,\psi)}_{(ab)} S^{(J,\psi)}_{(ab)\widehat{(p,\chi)}}+ 
\sum_{\widehat{(a,\xi)}} S^{(J,\psi)}_{(mn)\widehat{(a,\xi)}} T^{(J,\psi)}_{\widehat{(a,\xi)}} S^{(J,\psi)}_{\widehat{(a,\xi)}\widehat{(p,\chi)}}=\nonumber\\
&&=
\delta_{\psi,0}\,\sum_{\xi=0}^1 \sum_{a,\,Ja=a} S^J_{ma}\,S^J_{na}\, e^{2i\pi(2\,h_a-\frac{c}{12})}\,
C\,\frac{1}{2}\,S_{ap}\,e^{i\pi\xi}+\nonumber\\
&&+
\sum_{(ab)} \left(S^J_{ma}\,S^J_{nb}+(-1)^\psi S^J_{mb}\,S^J_{na}\right)\, e^{2i\pi(h_a +h_b-\frac{c}{12})}\,
\cdot
\left\{
\begin{array}{cl}
0 & {\,\,\rm\,\,} Ja=a\\
A\,S_{ap} & {\,\,\rm\,\,} Ja=b
\end{array}
\right. +\nonumber\\
&&+
\sum_{\xi=0}^1 \sum_{a,\,\widehat{(a,\xi)}\,\,{\rm f.p.of\,\,}(J,\psi)} A\,
S_{ma}\,e^{2i\pi(\frac{h_a}{2}+\frac{\xi}{2}-\frac{c}{48})}\,
B\,\frac{1}{2}\,e^{i\pi\hat{Q}_J(a)}\,P_{Ja,p}\,\,e^{i\pi(\xi+\chi)}\nonumber\,.
\end{eqnarray}
Now the first line vanishes because of $\sum_{\xi=0}^1 e^{i\pi\xi}=0$. The second line also vanishes since $S^J=0$ for orbit fields. In the third line we can use (\ref{P_(Jp,q)}). So we get
\begin{eqnarray}
&& \left(S^{(J,\psi)} T^{(J,\psi)} S^{(J,\psi)}\right)_{(mn)\widehat{(p,\chi)}}=\dots=
\nonumber\\
&&=
A\,B\,e^{i\pi\chi} \sum_{a,\,\widehat{(a,\xi)}\,\,{\rm f.p.of\,\,}(J,\psi)} 
S_{ma}\,\sqrt{T}_a\,\,e^{i\pi\hat{Q}_J(a)}\,P_{Ja,p}=\nonumber\\
&&=
A\,B\,e^{i\pi\chi}\, e^{i\pi h_J}\,\sum_l\left( \sum_{a,\,\widehat{(a,\xi)}\,\,{\rm f.p.of\,\,}(J,\psi)} 
S_{ma}\,T_a\,S_{al}\right) 
e^{2i\pi Q_J(l)}\,T^2_l\,S_{lp}\,\sqrt{T}_p\,.\nonumber
\end{eqnarray}
Let us look more in detail at the quantity within brackets. In order to extend the sum over all fields we need to use the projector (\ref{projector PI}):
\begin{eqnarray}
&& \sum_{a,\,\widehat{(a,\xi)}\,\,{\rm f.p.of\,\,}(J,\psi)} 
S_{ma}\,T_a\,S_{al}=
\sum_a \Pi(a)\, S_{ma}\,T_a\,S_{al}=\nonumber\\
&&=
\sum_a \frac{1}{2}\,\sum_{k=0}^1\,e^{2i\pi k(Q_J(a)-\frac{\psi}{2}-2h_J)} S_{ma}\,T_a\,S_{al}=\nonumber\\
&&=
\sum_{k=0}^1\,\frac{1}{2}\,e^{-2i\pi k(\frac{\psi}{2}+2h_J)}\sum_a S_{J^km,a}\,T_a\,S_{al}=\nonumber\\
&&=
\frac{1}{2}\,\sum_a \,S_{ma}\,T_a\,S_{al}+
(-1)^{\psi+4h_J}\,\frac{1}{2}\,\sum_a S_{ma}\,T_a\,S_{al}=\nonumber\\
&&=
\frac{1}{2}\,(STS)_{ml}+(-1)^{\psi+4h_J}\,\frac{1}{2}\,(STS)_{nl}\,.\nonumber
\end{eqnarray}
Going back to our main expression, we can interchange $STS=T^{-1}ST^{-1}$ and get
\begin{eqnarray}
&& \left(S^{(J,\psi)} T^{(J,\psi)} S^{(J,\psi)}\right)_{(mn)\widehat{(p,\chi)}}=\dots=
\nonumber\\
&&=
A\,B\,e^{i\pi\chi}\, e^{i\pi h_J}\,\frac{1}{2}\sum_l\,
[(STS)_{ml}+(-1)^{\psi+4h_J}\,(STS)_{nl}]\,
e^{2i\pi Q_J(l)}\,T^2_l\,S_{lp}\,\sqrt{T}_p=\nonumber\\
&&=
A\,B\,e^{i\pi\chi}\, e^{i\pi h_J}\,\frac{1}{2}\sum_l\,
[T^{-1}_m\,S_{ml}\,T^{-1}_l+(-1)^{\psi+4h_J}\,T^{-1}_n\,S_{nl}\,T^{-1}_l]\,
e^{2i\pi Q_J(l)}\,T^2_l\,S_{lp}\,\sqrt{T}_p=\nonumber\\
&&=
A\,B\,e^{i\pi\chi}\, e^{i\pi h_J}\,\frac{1}{2}\sum_l\,
[T^{-1}_m\,S_{Jm,l}\,T^{-1}_l+(-1)^{\psi+4h_J}\,T^{-1}_n\,S_{Jn,l}\,T^{-1}_l]\,
T^2_l\,S_{lp}\,\sqrt{T}_p=\nonumber\\
&&=
A\,B\,e^{i\pi\chi}\, e^{i\pi h_J}\,\frac{1}{2}\,
[T^{-1}_m\,(STS\sqrt{T})_{np}+(-1)^{\psi+4h_J}\,T^{-1}_n\,(STS\sqrt{T})_{mp}]=\nonumber\\
&&=
A\,B\,e^{i\pi\chi}\, e^{i\pi h_J}\,\frac{1}{2}\,
[T^{-1}_m\,T^{-1}_n\,S_{np}\,\sqrt{T}^{-1}_p+(-1)^{\psi+4h_J}\,T^{-1}_n\,T^{-1}_m\,S_{mp}\,\sqrt{T}^{-1}_p]=\nonumber\\
&&=
A\,B\,e^{i\pi\chi}\, e^{i\pi h_J}\,\frac{1}{2}\,T^{-1}_m\,T^{-1}_n\,
(\underbrace{S_{np}}_{(-1)^{\psi+4h_J}\,S_{mp}}+(-1)^{\psi+4h_J}\,S_{mp})\,\sqrt{T}^{-1}_p=\nonumber\\
&&=
A\,B\,e^{i\pi\chi}\, (-1)^{\psi+4h_J}\,e^{i\pi h_J}\,T^{-1}_m\,T^{-1}_n\,S_{mp}\,\sqrt{T}^{-1}_p\,.\nonumber
\end{eqnarray}
Hence
\begin{equation}
\left(S^{(J,\psi)} T^{(J,\psi)} S^{(J,\psi)}\right)_{(mn)\widehat{(p,\chi)}}=
\left({T^{(J,\psi)}}^{-1} S^{(J,\psi)} {T^{(J,\psi)}}^{-1}\right)_{(mn)\widehat{(p,\chi)}}\,,
\end{equation}
provided
\begin{equation}
\boxed{
B=(-1)^{\psi+4h_J}\,e^{-i\pi h_J}=(-1)^{\psi}\,e^{3i\pi h_J}}\,.
\end{equation}

\subsection{Summary of phase relations}
Before ending this appendix we think it is useful to summarize the phase relations that we have found in the calculation for modular invariance.

Combining the results from the twisted-twisted, diagonal-twisted and off-diagonal-twisted sectors, one can see that
\begin{eqnarray}
B   &=&(-1)^\psi\,e^{3 i\pi h_J} \nonumber\\
A^2 = C^2 &=&B\,e^{-i\pi h_J}=(-1)^\psi\,e^{2i\pi h_J}\nonumber
\end{eqnarray}
There is no phase information coming from the off-diagonal-off-diagonal sector. As far as the remaining phases are concerned, it is straightforward to show (even if we have not reported it explicitly in this appendix) that the phase of $S^{(J,\psi)}_{(mn)(pq)}$ and $S^{(J,\psi)}_{(i,\phi)(j,\chi)}$ are constrained to be equal to one, so that there is no additional phase in these matrix elements, while the square of the phase of $S^{(J,\psi)}_{(i,\phi)(mn)}$ must be one as well, so that one cannot fix the sign of this matrix entry by using modular invariance.

Let us remark a subtle point. From the expressions above we can see that $B$ is fixed while $A$ and $C$ are fixed up to a sign (we could in principle choose both signs for the square root). We have already remarked that the phase $C$ is relevant only for $\psi=0$, since for $\psi=1$ there are no diagonal fixed points and hence no matrix element $S^{(J,\psi)}_{(i,\phi)\widehat{(p,\chi)}}$.

\end{document}